\documentclass[twocolumn]{aastex631}

\usepackage[T1]{fontenc} 
\usepackage[utf8]{inputenc} 

\newcommand\tess{TESS}
\newcommand\gaia{\textit{Gaia}}
\newcommand\kms{$\textrm{km~s}^{-1}$}
\newcommand\ms{$\textrm{m~s}^{-1}$}
\newcommand\cms{$\textrm{cm~s}^{-1}$}
\newcommand\gcmcubed{$\textrm{g~cm}^{-3}$}

\newcommand\teff{T$_{\rm{eff}}$}

\newcommand{\unit}[1]{\ensuremath{\, \mathrm{#1}}} 
\newcommand\earthmass{$M_{\oplus}$}
\newcommand\earthradius{$R_{\oplus}$}
\newcommand\solmass{$M_{\odot}$}

\hypersetup{linkcolor=teal,citecolor=teal}

\graphicspath{{./}{figures/}}

\usepackage{inputenc}

\received{June 11, 2024}
\revised{August 16, 2024}
\accepted{August 20, 2024}
\submitjournal{AJ}

\begin{document}

\title{\textit{Searching for GEMS:} Characterizing  Six Giant Planets around Cool Dwarfs}

\author[0000-0001-8401-4300]{Shubham Kanodia}
\affil{Earth and Planets Laboratory, Carnegie Science, 5241 Broad Branch Road, NW, Washington, DC 20015, USA}

\author[0000-0002-5463-9980]{Arvind F.\ Gupta}
\affil{U.S. National Science Foundation National Optical-Infrared Astronomy Research Laboratory, 950 N. Cherry Ave., Tucson, AZ 85719, USA}

\author[0000-0003-4835-0619]{Caleb I. Ca\~nas}
\altaffiliation{NASA Postdoctoral Fellow}
\affiliation{NASA Goddard Space Flight Center, 8800 Greenbelt Road, Greenbelt, MD 20771, USA }

\author[0000-0002-8035-1032]{Lia Marta Bernab\`o}
\affiliation{Institute of Planetary Research, German Aerospace Center (DLR), Rutherfordstrasse 2, 12489 Berlin}

\author[0009-0006-7298-619X]{Varghese Reji}
\affil{Department of Astronomy and Astrophysics, Tata Institute of Fundamental Research, Homi Bhabha Road, Colaba, Mumbai 400005, India}

\author[0000-0002-7127-7643]{Te Han}
\affil{Department of Physics \& Astronomy, The University of California, Irvine, Irvine, CA 92697, USA}

\author[0000-0003-2404-2427]{Madison Brady}
\affil{Department of Astronomy \& Astrophysics, University of Chicago, Chicago, IL 60637, USA}

\author[0000-0003-4526-3747]{Andreas Seifahrt}
\affil{Department of Astronomy \& Astrophysics, University of Chicago, Chicago, IL 60637, USA}

\author[0000-0001-9662-3496]{William D. Cochran}
\affil{McDonald Observatory and Department of Astronomy, The University of Texas at Austin}
\affil{Center for Planetary Systems Habitability, The University of Texas at Austin}

\author[0000-0003-2535-3091]{Nidia Morrell}
\affil{Las Campanas Observatory, Carnegie Observatories, Casilla 601, La Serena, Chile}


\author[0000-0003-4508-2436]{Ritvik Basant}
\affil{Department of Astronomy \& Astrophysics, University of Chicago, Chicago, IL 60637, USA}

\author[0000-0003-4733-6532]{Jacob Bean}
\affil{Department of Astronomy \& Astrophysics, University of Chicago, Chicago, IL 60637, USA}

\author[0000-0003-4384-7220]{Chad F.\ Bender}
\affil{Steward Observatory, The University of Arizona, 933 N. Cherry Ave, Tucson, AZ 85721, USA}

\author[0000-0002-7564-6047]{Zo\"e L. de Beurs}\altaffiliation{NSF Graduate Research Fellow}
\altaffiliation{MIT Presidential Fellow, MIT Collamore-Rogers Fellow}
\affil{Department of Earth, Atmospheric and Planetary Sciences, Massachusetts Institute of Technology,  Cambridge,  MA 02139, USA}

\author[0000-0001-6637-5401]{Allyson~Bieryla} 
\affiliation{Center for Astrophysics ${\rm \mid}$ Harvard {\rm \&} Smithsonian, 60 Garden Street, Cambridge, MA 02138, USA}

\author{Alexina Birkholz}
\affil{Department of Physics \& Astronomy, University of Wyoming, Laramie, WY 82070, USA}

\author[0009-0003-1142-292X]{Nina Brown}
\affil{Department of Astronomy \& Astrophysics, University of Chicago, Chicago, IL 60637, USA}

\author[0009-0003-9699-1063]{Franklin Chapman}
\affiliation{Department of Physics \& Astronomy, University of Wyoming, Laramie, WY 82070, USA}

\author[0000-0002-5741-3047]{David~R.~Ciardi}
\affil{NASA Exoplanet Science Institute-Caltech/IPAC, Pasadena, CA 91125, USA}

\author[0000-0002-2361-5812]{Catherine A. Clark}
\affil{Jet Propulsion Laboratory, California Institute of Technology, Pasadena, CA 91109 USA}
\affil{NASA Exoplanet Science Institute, IPAC, California Institute of Technology, Pasadena, CA 91125 USA}

\author[0009-0003-1637-8315]{Ethan G. Cotter}
\affil{Department of Physics \& Astronomy, University of Wyoming, Laramie, WY 82070, USA}

\author[0000-0002-2144-0764]{Scott A. Diddams}
\affil{Electrical, Computer \& Energy Engineering, University of Colorado, 1111 Engineering Dr.,  Boulder, CO 80309, USA}
\affil{Department of Physics, University of Colorado, 2000 Colorado Avenue, Boulder, CO 80309, USA}

\author[0000-0003-1312-9391]{Samuel Halverson}
\affil{Jet Propulsion Laboratory, 4800 Oak Grove Drive, Pasadena, CA 91109, USA}

\author[0000-0002-6629-4182]{Suzanne Hawley}
\affil{Department of Astronomy, Box 351580, University of Washington, Seattle, WA 98195 USA}

\author[0000-0003-1263-8637]{Leslie Hebb}
\affiliation{Department of Physics, Hobart and William Smith Colleges, 300 Pulteney Street, Geneva,
NY, 14456, USA}

\author[0000-0002-5034-9476]{Rae Holcomb}
\affil{Department of Physics \& Astronomy, The University of California, Irvine, Irvine, CA 92697, USA}

\author[0000-0002-2532-2853]{Steve~B.~Howell}
\affil{NASA Ames Research Center, Moffett Field, CA 94035, USA}

\author[0000-0002-4475-4176]{Henry A. Kobulnicky}
\affil{Department of Physics \& Astronomy, University of Wyoming, Laramie, WY 82070, USA}

\author[0000-0001-7458-1176]{Adam F. Kowalski}
\affil{Department of Astrophysical and Planetary Sciences, University of Colorado Boulder, 2000 Colorado Ave,
Boulder, CO 80305, USA}
\affil{National Solar Observatory, University of Colorado Boulder, 3665 Discovery Drive, Boulder, CO 80303, USA}
\affil{Laboratory for Atmospheric and Space Physics, University of Colorado Boulder, 3665 Discovery Drive, Boulder, CO 80303, USA}

\author[0000-0002-2401-8411]{Alexander Larsen}
\affiliation{Department of Physics \& Astronomy, University of Wyoming, Laramie, WY 82070, USA}

\author[0000-0002-2990-7613]{Jessica Libby-Roberts}
\affil{Department of Astronomy \& Astrophysics, 525 Davey Laboratory, The Pennsylvania State University, University Park, PA 16802, USA}
\affil{Center for Exoplanets and Habitable Worlds, 525 Davey Laboratory, The Pennsylvania State University, University Park, PA 16802, USA}

\author[0000-0002-9082-6337]{Andrea S.J.\ Lin}
\affil{Department of Astronomy \& Astrophysics, 525 Davey Laboratory, The Pennsylvania State University, University Park, PA 16802, USA}
\affil{Center for Exoplanets and Habitable Worlds, 525 Davey Laboratory, The Pennsylvania State University, University Park, PA 16802, USA} 

\author[0000-0003-2527-1598]{Michael B. Lund}
\affil{NASA Exoplanet Science Institute-Caltech/IPAC, Pasadena, CA 91125, USA}

\author[0000-0002-4671-2957]{Rafael Luque}
\affil{Department of Astronomy \& Astrophysics, University of Chicago, Chicago, IL 60637, USA}

\author[0000-0002-0048-2586]{Andrew Monson}
\affil{Steward Observatory, The University of Arizona, 933 N.\ Cherry Avenue, Tucson, AZ 85721, USA}

\author[0000-0001-8720-5612]{Joe P.\ Ninan}
\affil{Department of Astronomy and Astrophysics, Tata Institute of Fundamental Research, Homi Bhabha Road, Colaba, Mumbai 400005, India}

\author[0000-0001-9307-8170]{Brock A. Parker}
\affil{Steward Observatory, The University of Arizona, 933 N. Cherry Ave, Tucson, AZ 85721, USA}

\author{Nishka Patel}
\affil{Maggie L.\ Walker Governor's School, Richmond, VA, 23220, USA}

\author{Michael Rodruck}
\affil{Department of Physics, Engineering, and Astrophysics, Randolph-Macon College, Ashland, VA 23005, USA}

\author[0009-0006-7023-1199]{Gabrielle Ross}
\affil{Department of Astrophysical Sciences, Princeton University, 4 Ivy Lane, Princeton, NJ 08540, USA}

\author[0000-0001-8127-5775]{Arpita Roy}
\affiliation{Astrophysics \& Space Institute, Schmidt Sciences, New York, NY 10011, USA}

\author[0000-0002-4046-987X]{Christian Schwab}
\affil{School of Mathematical and Physical Sciences, Macquarie University, Balaclava Road, North Ryde, NSW 2109, Australia}

\author[0000-0001-7409-5688]{Guðmundur Stefánsson} 
\affil{Anton Pannekoek Institute for Astronomy, University of Amsterdam, Science Park 904, 1098 XH Amsterdam, The Netherlands} 

\author{Aubrie Thoms}
\affil{Department of Physics, Engineering, and Astrophysics, Randolph-Macon College, Ashland, VA 23005, USA}

\author[0000-0001-7246-5438]{Andrew Vanderburg}
\affil{Department of Physics and Kavli Institute for Astrophysics and Space Research, Massachusetts Institute of Technology, Cambridge, MA 02139, USA}

\shortauthors{Kanodia et al. 2024}

\correspondingauthor{Shubham Kanodia}
\email{skanodia@carnegiescience.edu}

\begin{abstract}
Transiting giant exoplanets around M-dwarf stars (GEMS) are rare, owing to the low-mass host stars. However, the all-sky coverage of TESS has enabled the detection of an increasingly large number of them to enable statistical surveys like the \textit{Searching for GEMS} survey. As part of this endeavour, we describe the observations of six transiting giant planets, which includes precise mass measurements for two GEMS (K2-419Ab, TOI-6034b) and statistical validation for four systems, which includes validation and mass upper limits for three of them (TOI-5218b, TOI-5616b, TOI-5634Ab), while the fourth one --- TOI-5414b is classified as a `likely planet'. Our observations include radial velocities from the Habitable-zone Planet Finder on the Hobby-Eberly Telescope, and MAROON-X on Gemini-North, along with photometry and high-contrast imaging from multiple ground-based facilities. In addition to TESS photometry, K2-419Ab was also observed and statistically validated as part of the K2 mission in Campaigns 5 and 18, which provides precise orbital and planetary constraints despite the faint host star and long orbital period of $\sim 20.4$ days. With an equilibrium temperature of only 380 K, K2-419Ab is one of the coolest known well-characterized transiting planets. TOI-6034 has a late F-type companion about 40\arcsec~away, making it the first GEMS host star to have an earlier main-sequence binary companion. These confirmations add to the existing small sample of confirmed transiting GEMS.
\end{abstract}


\section{Introduction} \label{sec:intro}
Transiting giant exoplanets around M-dwarf stars (GEMS) with radii $>$ 8 \earthradius{} are rare outcomes of planet formation, with preliminary estimates of occurrence for the short-period ones ($<$ 10 days) at $\sim$ 0.1\% \citep{gan_occurrence_2023, bryant_occurrence_2023}. The sample of these planets (3 pre-TESS) has been bolstered by recent detections from NASA's TESS mission. Despite this, the current size of the sample ($\sim$ 20) precludes robust statistical studies. Therefore, we have started the \textit{Searching for GEMS} survey to provide for occurrence rates for short-period transiting GEMS across a well-characterized homogeneous sample of a million M-dwarfs observed by TESS. In addition, this survey and similar community efforts \citep[e.g. MANGOS;][]{triaud_m_2023} will provide mass measurements for a large number of transiting GEMS. These mass measurements will enable a comparison of the bulk properties of these objects with their FGK short-period analogues --- hot Jupiters.  

GEMS likely form in an extremely mass-starved regime \citep{kanodia_toi-5205b_2023, delamer_toi-4201_2024}, which raises the questions as to whether this influences their present-day bulk-properties or could lead to a different host star metallicity dependence \citep{johnson_metal_2009}.  Beyond bulk-comparisons, precise characterization of these planets enables subsequent atmospheric observations with JWST (Cycle 2 GO 3171, 3731, 4227, Cycle 3 GO 5863), as well as 3D orbital measurements through the Rossiter-McLaughlin effect \citep{rossiter_detection_1924, mclaughlin_results_1924, albrecht_stellar_2022}. 

Understanding the process of giant planet formation is crucial to explain the architecture of exoplanetary systems. Studies have shown that the presence of gas giant planets affects the formation and evolution of smaller terrestrial planets both within the Solar System \citep{raymond_building_2009, brasser_analysis_2016}, and in extra-solar systems \citep{2021ApJ...920L...1M}.  In particular, close-in giant planets like transiting GEMS and hot Jupiters have confounded traditional planet formation theories \citep{dawson_origins_2018}, necessitating formation farther out and then migration inwards \citep{alibert_models_2005}. Extending these giant planet samples to include cool dwarfs can enable comparisons across the stellar mass axis as a test of planet formation (and migration) models by utilizing this dependency.

In this study we perform detailed characterization of six TESS objects of interest (TOIs), where we provide mass measurements for two GEMS (K2-419Ab and TOI-6034b), along with upper mass limits for four giant planets, of which two orbit M-dwarfs (TOI-5616b and TOI-5634Ab) while two orbit late K-dwarfs (TOI-5218b and TOI-5414b). The paper is structured as follows: in Section \ref{sec:observations} we describe all the observations utilized in this work to validate and characterize the six planets. In Section \ref{sec:stellar} we detail the stellar characterization performed, while Section \ref{sec:joint} describes the procedure followed to jointly model the data and derive planetary parameters. In Section \ref{sec:discussion} we discuss this work in the context of the existing planet sample and contextualize its importance before concluding in Section \ref{sec:conclusion}.

\vspace{1 cm}

\section{Observations}\label{sec:observations}

\subsection{K2-419A}
\subsubsection{Photometry}\label{sec:5176phot}

\begin{figure*}[!t]
\begin{center}
\fig{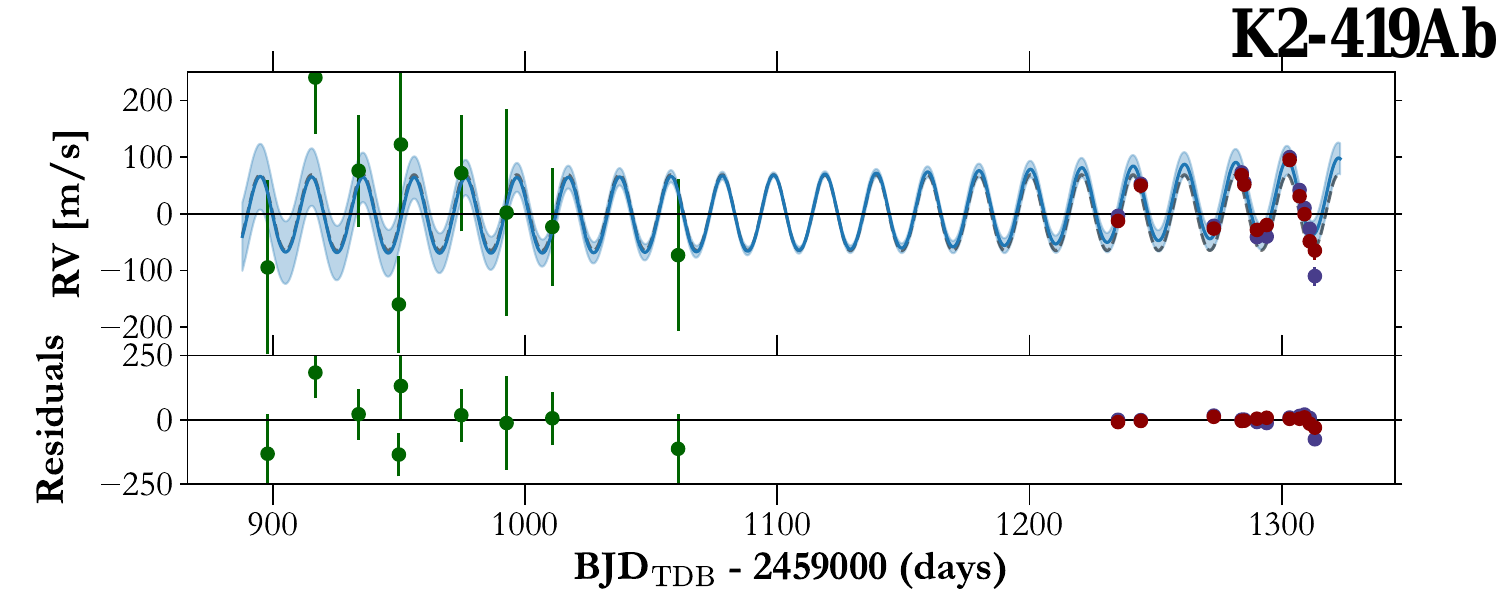}{0.62\textwidth}{}
\fig{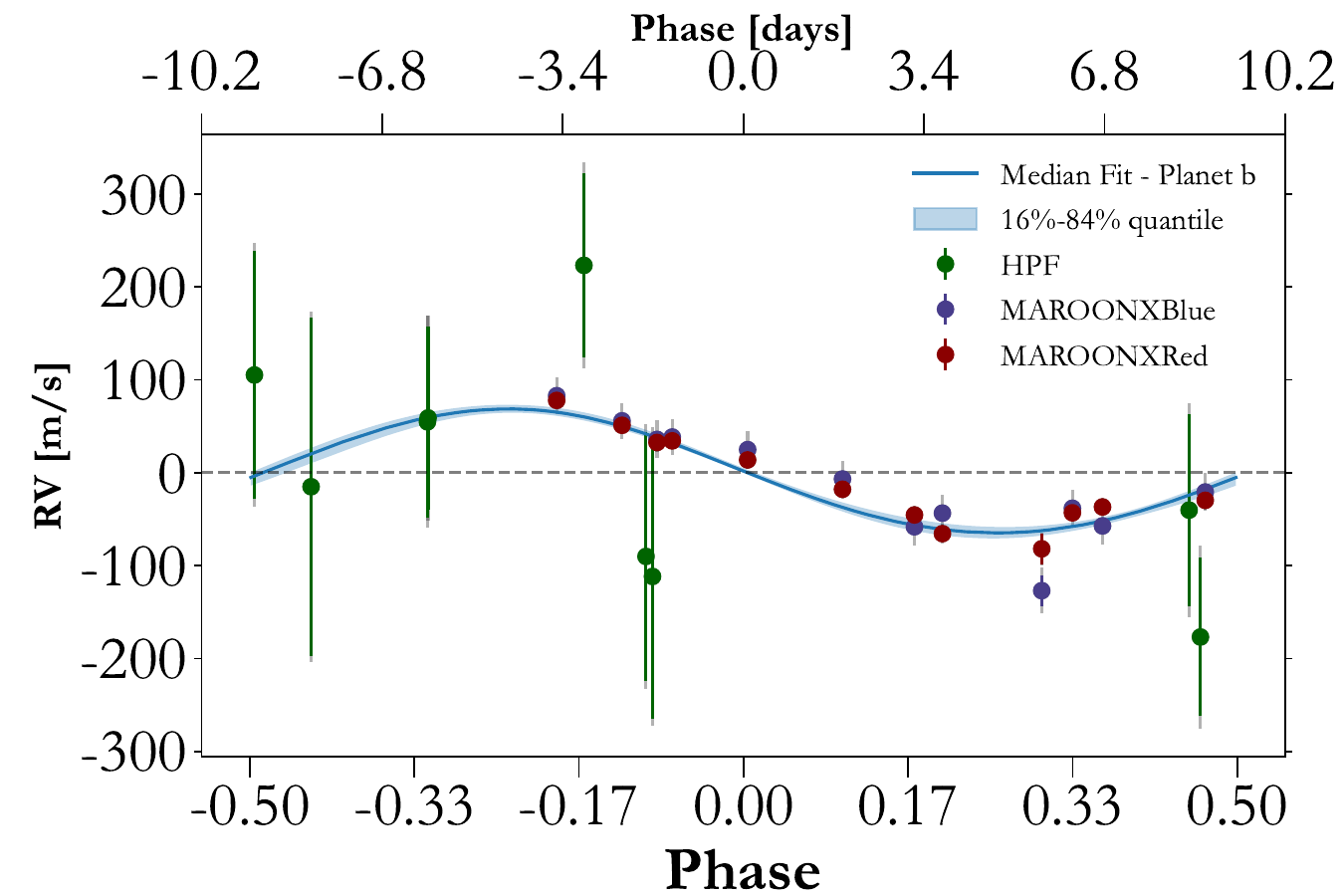}{0.36\textwidth}{}
\fig{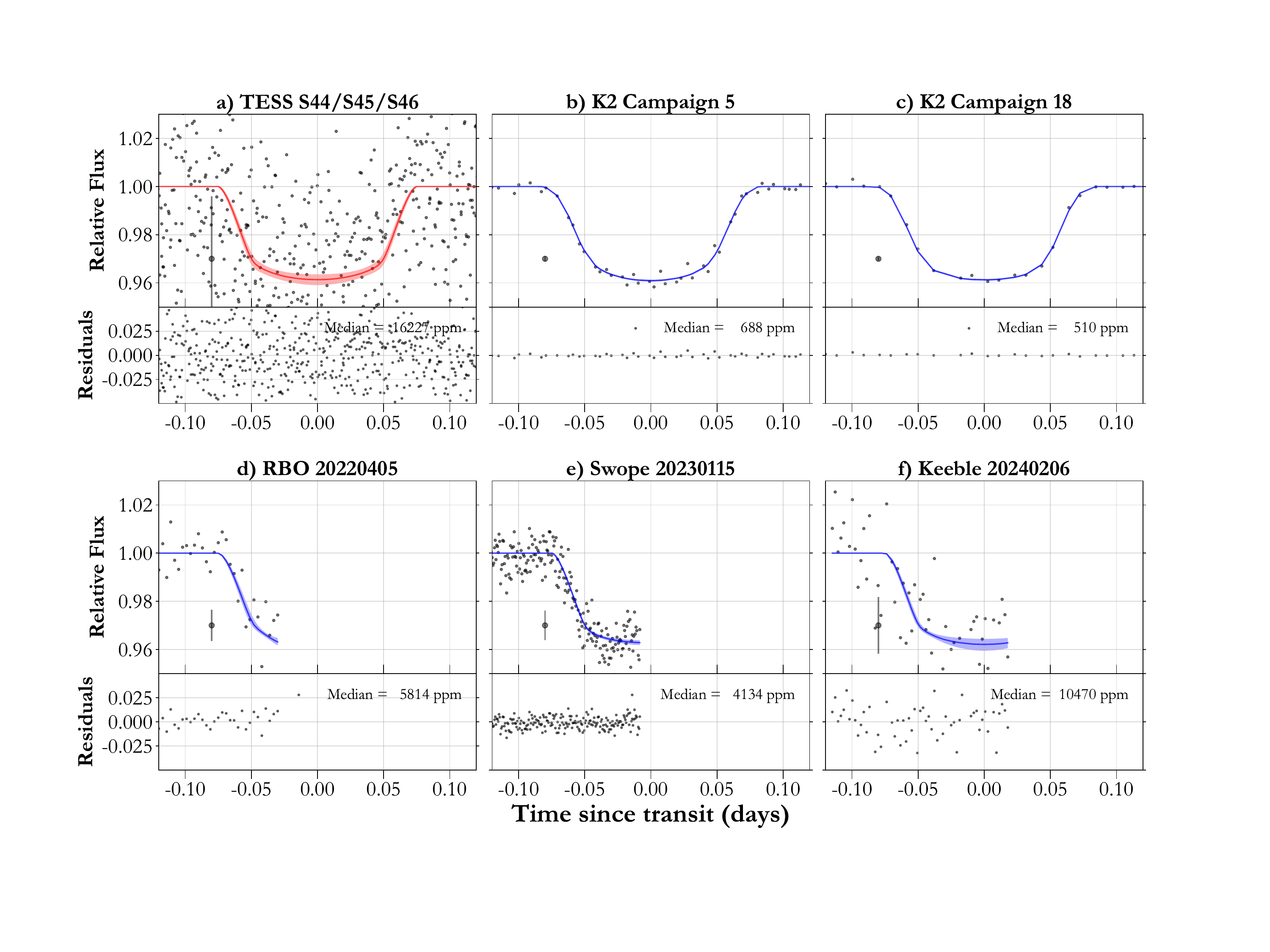}{0.95\textwidth}{}
\end{center}
\vspace{-2 cm}
\caption{\small \textbf{Top Left)} RV time series for K2-419A, where the dashed line denotes the planet model, whereas the blue line and shaded region depicts the best model fit (and 16 -- 84\% uncertainties) after including the RV offsets and trends. \textbf{Top Right)} RVs phase folded at the best-fit period, including the model and uncertainty. The grey errorbars behind the points show the errors when including the RV jitter that is estimated during the joint fit. \textbf{Bottom)} Phase folded photometry for K2-419A, including TESS (shown in red), where the dilution term is allowed to float owing to its large pixel size. The model and uncertainties follow the usual definitions. The TESS and K2 data (a,b,c) are shown after subtracting the median trend as estimated after masking the planet transits.  We also include the representative median statistical uncertainty at -0.07 days. }\label{fig:Data5176}
\end{figure*}

\textbf{K2:} K2-419A (EPIC-211509553A) was observed by the \textit{Kepler} spacecraft in long cadence mode (thirty-minute cadence) as part of Campaigns 5 and 18 during the K2 mission (\autoref{tab:photometric}). The star was originally identified as a candidate planet host by \cite{dressing2017} and statistically validated using ground-based spectra and high-contrast imaging \citep{livingston2018,petigurak2}. For Campaign 18 we used the \texttt{EVEREST} pipeline \citep{luger_everest_2016, luger_update_2018} to correct for photometric variations seen in the K2 photometry due to imperfect pointing of the spacecraft with only two functioning reaction wheels. We exclude all points more than one transit duration away from each mid-transit with non-zero K2 data quality flags. For Campaign 5, we find that the \texttt{EVEREST} light curves are over-correcting the systematics outside of transit, and instead use the Campaign 5 reduced light curves from \cite{dressing2017}, which were produced following the methodology of \cite{vanderburg_planetary_2016}. The detrended phase-folded K2 photometry (see Section \ref{sec:joint} regarding detrending) is shown in \autoref{fig:Data5176}.

\begin{longtable*}{ccccc}
\caption{Summary of ground- and space- based photometric follow up. Instruments marked with a * are excluded from the joint analysis. We include a \texttt{tar} file containing individual \texttt{npy} binary pickled \texttt{python (v3.9)} dictionaries for each TOI created using \texttt{numpy (v1.20)}. These dictionaries contain the photometry time series (including flux errors) for each instrument used in this analysis, as well as the model used for detrending where applicable. This can also be made available in other formats upon reasonable request}. \label{tab:photometric} \\
\hline \hline
\textbf{Date} & \textbf{Instrument} & \textbf{Filter} & \textbf{Exposure Time} & \textbf{Median PSF} \\
\textbf{UTC} & & & \textbf{Time (s)} & \textbf{FWHM (\arcsec)} \\
\hline
\endfirsthead
\caption{Summary of ground- and space- based photometric follow up (continued from previous page). } \\
\hline \hline
\textbf{Date} & \textbf{Instrument} & \textbf{Filter} & \textbf{Exposure Time} & \textbf{Median PSF} \\
\textbf{UTC} & & & \textbf{Time (s)} & \textbf{FWHM (\arcsec)} \\
\hline
\endhead
\multicolumn{5}{l}{\hspace{-0.2cm}\textbf{K2-419A}}  \\
2015 Apr 27 -- 2015 Jul 10 & K2/C5 & $Kp$ & 1800 & ... \\
2018 May 12 -- 2018 Jul 02 & K2/C18 & $Kp$ & 1800 & ... \\
2021 Oct 12 -- 2021 Nov 06 & TESS/S44 & $T$ & 120 & 39.5  \\
2021 Nov 06 -- 2021 Dec 02 & TESS/S45 & $T$ & 120 & 39.5  \\
2021 Dec 02 -- 2021 Dec 30 & TESS/S46 & $T$ & 120 & 39.5  \\
2022 Apr 06 & 0.6 m RBO & Bessell I & 240 & 1.58 \\
2023 Jan 16 & 1.0 m Swope & SDSS \textit{i'} & 35 & 3.67   \\
2024 Feb 07 & 0.4 m Keeble &  Cousins I & 180 & 3.51 \\
\multicolumn{5}{l}{\hspace{-0.2cm}\textbf{TOI-6034}}  \\
2019 Sep 11 -- 2019 Oct 07 & TESS/S16 & $T$ & 1800 & 39.5  \\
2019 Oct 07 -- 2019 Nov 02 & TESS/S17 & $T$ & 1800 & 39.5  \\
2019 Nov 02 -- 2019 Nov 27 & TESS/S18 & $T$ & 1800 & 39.5  \\
2020 Apr 16 -- 2020 May 13 & TESS/S24* & $T$ & 1800 & 39.5  \\
2020 May 13 -- 2020 Jun 08 & TESS/S25 & $T$ & 1800 & 39.5  \\
2022 May 18 -- 2022 Jun 13 & TESS/S52* & $T$ & 200 & 39.5  \\
2022 Sep 01 -- 2022 Sep 30 & TESS/S56 & $T$ & 200 & 39.5  \\
2022 Sep 30 -- 2022 Oct 29 & TESS/S57 & $T$ & 200 & 39.5  \\
2022 Oct 29 -- 2022 Nov 26 & TESS/S58* & $T$ & 200 & 39.5  \\
2023 Jun 20 & 0.6 m RBO & Bessell I & 240 & 1.66 \\
2023 Aug 08 & 0.6 m RBO & Bessell I & 240 & 1.44 \\
2023 Sep 08 & 0.6 m RBO & Bessell I & 240 & 1.51 \\
2023 Oct 09 & 0.4 m Keeble & Cousins I  & 120 & 5.6 \\
2018 Mar -- 2022 Oct & ZTF* & ZTF \textit{g} & 30 & 2.5 \\
2018 May -- 2023 Apr & ZTF* & ZTF \textit{r} & 30 & 2.5 \\
\multicolumn{5}{l}{\hspace{-0.2cm}\textbf{TOI-5218}}  \\
2019 Jul 18 -- 2019 Aug 14 & TESS/S14 & $T$ & 1800 & 39.5  \\
2019 Aug 15 -- 2019 Sep 10 & TESS/S15 & $T$ & 1800 & 39.5  \\
2019 Sep 11 -- 2019 Oct 07 & TESS/S16 & $T$ & 1800 & 39.5  \\
2019 Oct 07 -- 2019 Nov 02 & TESS/S17 & $T$ & 1800 & 39.5  \\
2019 Nov 02 -- 2019 Nov 27 & TESS/S18 & $T$ & 1800 & 39.5  \\
2019 Nov 28 -- 2019 Dec 23 & TESS/S19 & $T$ & 1800 & 39.5  \\
2019 Dec 24 -- 2020 Jan 20 & TESS/S20 & $T$ & 1800 & 39.5  \\
2020 Jan 21 -- 2020 Feb 18 & TESS/S21* & $T$ & 1800 & 39.5  \\
2020 Feb 19 -- 2020 Mar 17 & TESS/S22 & $T$ & 1800 & 39.5  \\
2020 Mar 19 -- 2020 Apr 15 & TESS/S23 & $T$ & 1800 & 39.5  \\
2020 Apr 16 -- 2020 May 12 & TESS/S24 & $T$ & 1800 & 39.5  \\
2020 May 14 -- 2020 Jun 08 & TESS/S25 & $T$ & 1800 & 39.5  \\
2020 Jun 09 -- 2020 Jul 04 & TESS/S26 & $T$ & 1800 & 39.5  \\
2021 Jun 24 -- 2021 Jul 23 & TESS/S40 & $T$ & 600 & 39.5  \\
2021 Jul 23 -- 2021 Aug 20 & TESS/S41 & $T$ & 600 & 39.5  \\
2021 Dec 30 -- 2022 Jan 28 & TESS/S47* & $T$ & 600 & 39.5  \\
2022 Jan 28 -- 2022 Feb 26 & TESS/S48* & $T$ & 600 & 39.5  \\
2022 Feb 26 -- 2022 Mar 26 & TESS/S49* & $T$ & 600 & 39.5  \\
2022 Mar 26 -- 2022 Apr 22 & TESS/S50* & $T$ & 600 & 39.5  \\
2022 Apr 22 -- 2022 May 18 & TESS/S51* & $T$ & 600 & 39.5  \\
2022 May 18 -- 2022 Jun 13 & TESS/S52* & $T$ & 600 & 39.5  \\
2022 Jun 13 -- 2022 Jul 09 & TESS/S53* & $T$ & 600 & 39.5  \\
2022 Jul 09 -- 2022 Aug 05 & TESS/S54* & $T$ & 600 & 39.5  \\
2022 Aug 05 -- 2022 Sep 01 & TESS/S55* & $T$ & 600 & 39.5  \\
2022 Sep 01 -- 2022 Sep 30 & TESS/S56* & $T$ & 200 & 39.5  \\
2022 Sep 30 -- 2022 Oct 29 & TESS/S57* & $T$ & 200 & 39.5  \\
2022 Oct 29 -- 2022 Nov 26 & TESS/S58* & $T$ & 200 & 39.5  \\
2022 Nov 26 -- 2022 Dec 23 & TESS/S59* & $T$ & 200 & 39.5  \\
2022 Dec 23 -- 2023 Jan 18 & TESS/S60* & $T$ & 200 & 39.5  \\
2022 May 14 & 3.5 m APO / ARCTIC & SDSS \textit{i'} & 20 & 2.8  \\
2022 Jul 13 & 3.5 m APO / ARCTIC & SDSS \textit{i'} & 20 & 2.8  \\
\multicolumn{5}{l}{\hspace{-0.2cm}\textbf{TOI-5414}}  \\
2019 Dec 24 -- 2020 Jan 20 & TESS/S20 & $T$ & 1800 & 39.5  \\
2021 Oct 12 -- 2021 Nov 06 & TESS/S44 & $T$ & 600 & 39.5  \\
2021 Nov 06 -- 2021 Dec 02 & TESS/S45 & $T$ & 600 & 39.5  \\
2021 Dec 30 -- 2022 Jan 28  & TESS/S47 & $T$ & 600 & 39.5  \\
2022 Nov 12 & 0.4 m LCOGT / Teide & $i^\prime$ & 600 & 2.64 \\
\multicolumn{5}{l}{\hspace{-0.2cm}\textbf{TOI-5616}}  \\
2020 Feb 18 -- 2020 Mar 18 & TESS/S22 & $T$ & 1800 & 39.5  \\
2022 Jan 28 -- 2022 Feb 26 & TESS/S48 & $T$ & 600 & 39.5  \\
2022 Feb 26 -- 2022 Mar 26 & TESS/S49 & $T$ & 600 & 39.5  \\
2022 May 31 & 1.2 m FLWO / KeplerCam & $i^\prime$ & 300 & 1.9 \\
2018 Mar -- 2023 May & ZTF* & ZTF \textit{g} & 30 & 0.6 \\
2018 Apr -- 2023 Apr & ZTF* & ZTF \textit{r} & 30 & 0.6 \\
\multicolumn{5}{l}{\hspace{-0.2cm}\textbf{TOI-5634A}}  \\
2020 Feb 18 -- 2020 Mar 18 & TESS/S22 & $T$ & 1800 & 39.5  \\
2022 Feb 26 -- 2022 Mar 26 & TESS/S49 & $T$ & 600 & 39.5  \\
2023 Jan 05 & 1.2 m FLWO / KeplerCam & $i^\prime$ & 300 & 2.1 \\
2023 Mar 20 & 0.3 m LCRO & $i^\prime$ & 420 & 1.3 \\
\hline
\end{longtable*}

\textbf{TESS / PDCSAP :} K2-419A (TOI-5176, TIC 437054764, 2MASS J09000476+1316258) was observed by \tess{} in Sectors 44, 45 and 46 in two minute cadence (\autoref{tab:photometric}). The data was processed by the \tess{} science processing pipeline \citep{smith_kepler_2012, stumpe_kepler_2012, stumpe_multiscale_2014,  jenkins_tess_2016}, and we used the pre-search data conditioned time-series light curves (PDCSAP) for our analysis that were available at the Mikulski Archive for Space Telescopes (MAST). We exclude points marked as anomalous by the \tess{} data quality flags. The detrended phase-folded photometry for all three TESS sectors is shown in \autoref{fig:Data5176}.


\textbf{0.6 m RBO :}  We used the 0.6 m telescope at the Red Buttes Observatory (RBO) in Wyoming \citep{kasper_remote_2016} to obtain a transit for K2-419Ab on 2022 April 6. This is an \textit{f}/8.43 Ritchey-Chrétien Cassegrain constructed by DFM Engineering, Inc. The observations were performed in Bessell I band, with 2$\times$2 on-chip binning and a binned plate scale of 0.73\arcsec. The observation is summarized in \autoref{tab:photometric}. The astrometry for the frames was verified using the \texttt{astrometry.net} package \citep{hogg_automated_2008}, while the aperture photometry was performed using the \texttt{photutils} package  \citep{bradley_astropyphotutils_2020} at the proper motion corrected locations for all \textit{Gaia} sources $G <$ 17 mag. Flat, bias and dark corrections were performed using \texttt{astropy} \citep{astropy_collaboration_astropy_2018}.  The exposure midpoints are converted to BJD$_\mathrm{TDB}$ using \texttt{barycorrpy} \citep{kanodia_python_2018}, following the recommendations and algorithms from \citep{eastman_achieving_2010}. The flux from the target star is compared to the sum of the fluxes for selected reference stars, which are present in all the frames, have no nearby companions in the aperture and are not known variables. The final apertures are chosen to minimize the scatter in the data, while the reference stars are manually chosen to yield consistent results across all the defined apertures. 

\textbf{1.0 m Swope telescope -- } We obtained a transit of K2-419A using the 1.0 m Henrietta Swope Telescope at the Las Campanas Observatory in Chile on 2023 January 16. The Swope telescope is an f/7 Ritchey-Chrétien equipped with the E2V 4K$\times$4K CCD. We observed through a Sloan $i$ filter. The detector was read in 2$\times$2 binning mode, which yields a pixel scale of 0.87\arcsec. Only the third quadrant (C3) was used for these observations resulting in a FOV of 14.8$\times$14.9 arcminutes. The telescope was defocussed in order to achieve $\sim$4\arcsec\ FWHM on the images. The target was observed continuously for almost six hours with individual exposures of 35~s. The observation is summarized in \autoref{tab:photometric}, with data-reduction and relative aperture photometry performed using \texttt{AstroImageJ} \citep[\texttt{AIJ};][]{collins_astroimagej_2017}. 

\textbf{0.4 m Keeble :} We used the 0.4 m telescope at the Randolph-Macon College's Keeble Observatory in Ashland, Virginia to observe a transit for K2-419A on UTC 2024 February 7. The Keeble telescope is an f/8 Ritchey–Chrétien manufactured by ASA. We used an SBIG ST-10XMEI CCD with a Kron/Cousins I-Band filter. All observations were binned in 2 $\times$ 2 mode with a plate scale of 0.88\arcsec. Our observations for both targets are summarized in \autoref{tab:photometric}. The CCD is characterized by a gain of 1.3 e$^-$/ADU. Data reduction and photometric analysis was performed with \texttt{AIJ}.


\begin{figure*}[!t]
    \fig{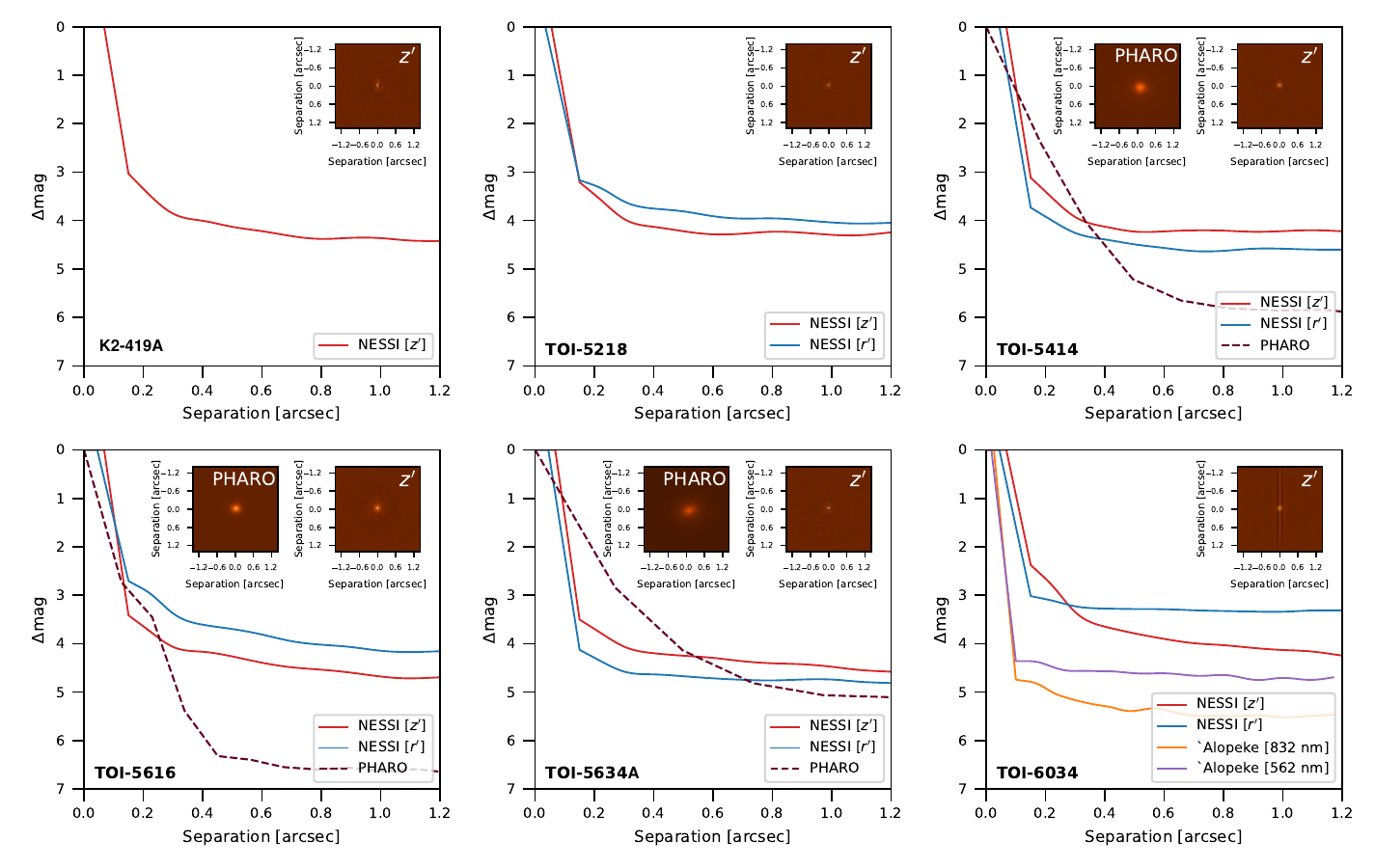}{\textwidth}{}
\caption{\small High contrast imaging data for all targets. In each panel, we show the achieved $5\sigma$ contrast limit as a function of angular separation from the host star for each high contrast data set. We also plot 1.4\arcsec\ $\times$ 1.4\arcsec\ postage stamps of the reconstructed $z'$-band NESSI images (right-hand inset in each panel)  for all targets and the AO images (left-hand insets) for TOI-5414, TOI-5616, and TOI-5634A and TOI-6034.}\label{fig:HighContrast}
\end{figure*}

\subsubsection{Radial Velocity}\label{sec:5176RVs}

\textbf{Habitable-zone Planet Finder (HPF) -- } We obtained 9 visits of K2-419A with the near-infrared precision spectrograph, HPF \citep{mahadevan_habitable-zone_2012, mahadevan_habitable-zone_2014}, at the 10-m Hobby-Eberly Telescope \citep[HET;][]{ramsey_progress_1988}. HPF is an environmentally stabilized spectrograph \citep{stefansson_versatile_2016}, with a stable fiber-fed illumination \citep{kanodia_overview_2018}. The data is processed using the algorithms described in the \texttt{HxRGproc} package \citep{ninan_habitable-zone_2018}, with the barycentric correction being performed using \texttt{barycorrpy} \citep{kanodia_python_2018} based on the algorithms from \citep{wright_barycentric_2014}. Similar to previous observations of faint TOIs with HPF, we do not use the simultaneous calibration using the laser-frequency comb \citep[LFC; ][]{metcalf_stellar_2019} due to concerns about scattered light from the comb, but instead interpolate the instrument wavelength solution from LFC observations taken at the beginning and end of the night, which has been shown to have a precision of $\sim30$ \cms{} per observation \citep{stefansson_sub-neptune-sized_2020}. We calculate RVs using a version of the template-matching algorithm \citep[e.g.,][]{anglada-escude_harps-terra_2012}, called \texttt{SERVAL} \citep{zechmeister_spectrum_2018}, which has been modified for use with HPF spectra \citep{stefansson_sub-neptune-sized_2020, stefansson_neptune-mass_2023}. The HPF data have a median per-pixel signal-to-noise (S/N) per unbinned exposure (of 969 s) at 1070 nm of 18. Each of our 9 visits consisted of 2 exposures of 969 seconds each, and were binned together. The HPF RVs for K2-419A are listed in \autoref{tab:rvs}. Given that K2-419A is a faint early M-dwarf ($J \sim 13.36$), our HPF RVs are insufficient to provide a mass measurement.

\textbf{MAROON-X -- } We obtained 12 visits of K2-419A with MAROON-X \citep{seifahrt_development_2016, seifahrt_-sky_2020, seifahrt_maroon-x_2022} at Gemini-N as part of program ID GN-2023B-Q-104 (PI: Kanodia). MAROON-X is a red-optical fiber-fed precise radial velocity spectrograph spanning 500 -- 920 nm across two arms (blue; 500 -- 670 nm and red; 650 -- 920 nm) with a resolving power of $\sim$ 85,000. The wavelength solution is obtained using a Fabry-Perot etalon. Each visit consisted of a 30 minute exposure, with the 2D spectra reduced by the MAROON-X team using a custom pipeline. The RVs were estimated for each arm separately using a modified version of \texttt{SERVAL} framework \citep{zechmeister_spectrum_2018}. We treat each arm as a separate instrument, and allow for different RV offsets and jitter terms while performing a joint fit (Section \ref{sec:joint}). Even though our data were taken over two separate observing runs, we do not treat them as separate since the known inter-run RV offsets are known to be $\lesssim$ 2 \ms{} \citep{seifahrt_maroon-x_2022}, which is much lower than our single-visit RV uncertainty of $\sim$ 10 \ms{}. The MAROON-X blue and red RVs are included in \autoref{tab:rvs}, and have a median S/N per pixel in the blue and red of 18 and 29 respectively.

\begin{deluxetable}{ccccc}
\tablecaption{RVs for the TOIs included in this paper. We show a few rows of the full table for representation purposes, while the rest of the RVs are included in  a machine readable file included with the manuscript. \label{tab:rvs}}
\tablehead{\colhead{System} & \colhead{$\unit{BJD_{TDB}}$}  &  \colhead{RV}   & \colhead{$\sigma$} & \colhead{Instrument}  \\
           \colhead{} & \colhead{days}   &  \colhead{\ms{}} & \colhead{\ms{}}}
\startdata
K2-419A & 2459897.92814 & -931.16 & 153.49 & HPF	 \\ 
K2-419A & 2459916.87028 & -596.19 & 98.95 & HPF \\ 
K2-419A & 2459934.00802 & -760.52 & 98.52 & HPF \\ 
K2-419A & 2459949.96576 & -996.36 & 85.19 & HPF \\ 
K2-419A & 2459950.77949 & -714.25 & 133.49 & HPF \\ 
... & ... & ... & ... & ... \\
\enddata
\end{deluxetable}

\subsubsection{High-contrast Imaging}\label{sec:5176HC}

To search for potential background sources, K2-419A was observed with the NN-Explore Exoplanet Stellar Speckle Imager \citep[NESSI;][]{scott_nn-explore_2018} on the WIYN\footnote{The WIYN Observatory is a joint facility of the NSF's National Optical-Infrared Astronomy Research Laboratory, Indiana University, the University of Wisconsin-Madison, Pennsylvania State University, Purdue University and Princeton University.} 3.5m telescope at Kitt Peak National Observatory. Speckle data were taken in the SDSS \textit{z}' filter using the red NESSI camera on the night of 2022 April 18, and the diffraction-limited frames were then used to reconstruct a high contrast image following the procedures described in \citet{howell_speckle_2011}. The reconstructed images and $5\sigma$ contrast limits are shown in \autoref{fig:HighContrast}. These data rule out the presence of any nearby sources brighter than $\Delta z' = 3.3$ mag at 0.2\arcsec\ and $\Delta z' = 4.4$ mag at 1.2\arcsec.

\subsection{TOI-6034}
\subsubsection{Photometry}\label{sec:6034phot}

\begin{figure*}[!t]
\begin{center}
\fig{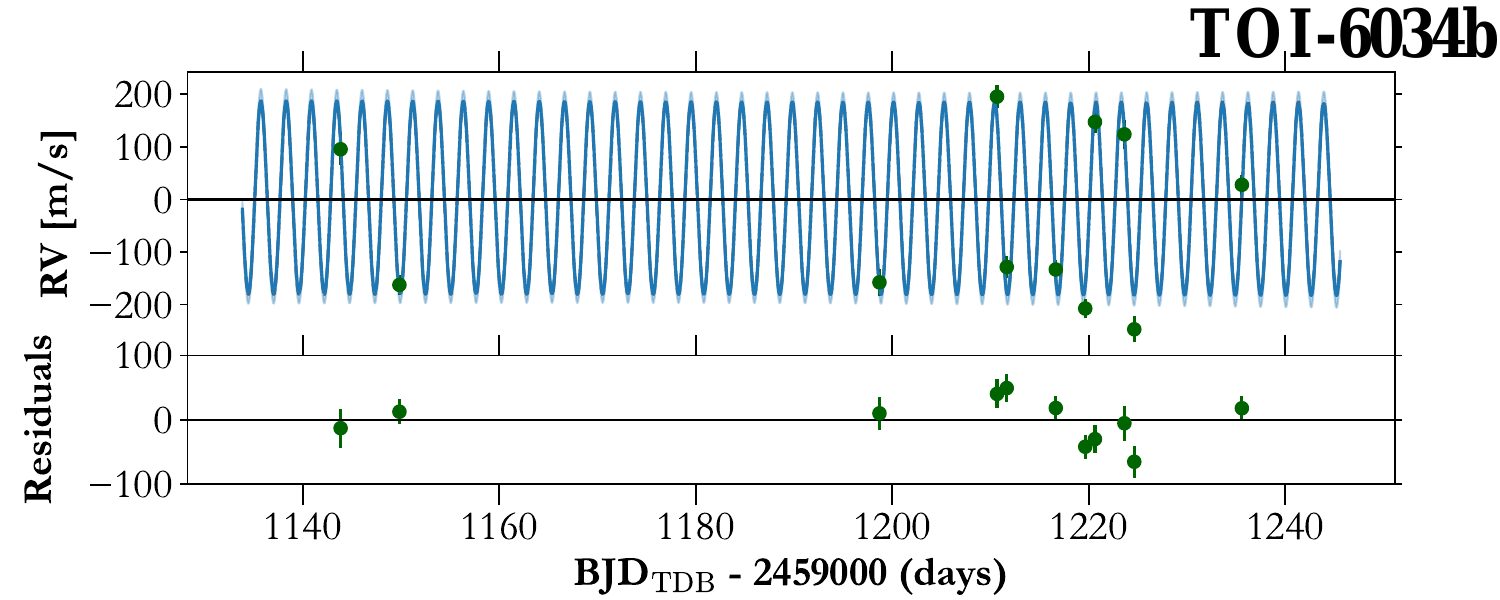}{0.62\textwidth}{}
\fig{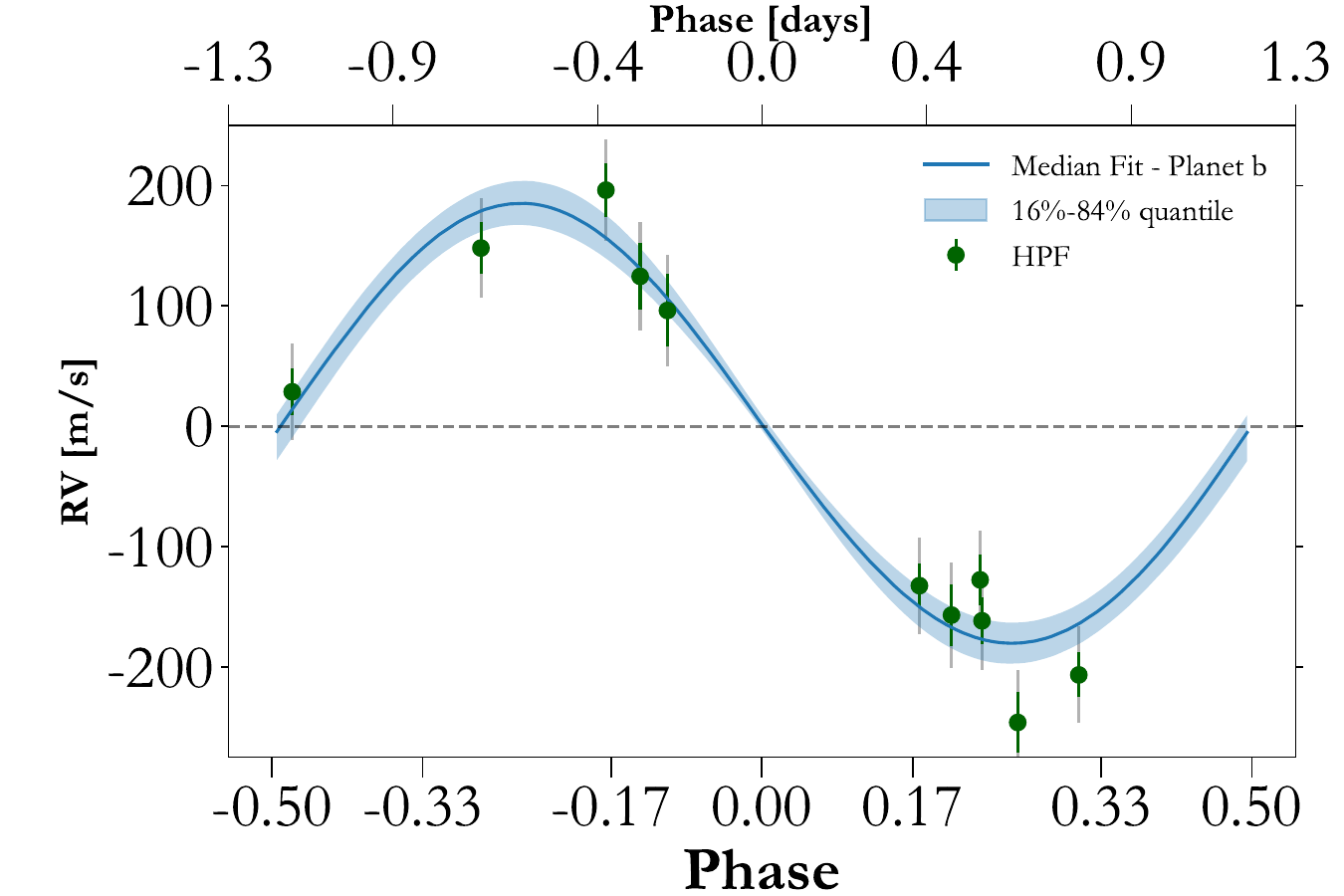}{0.36\textwidth}{}
\fig{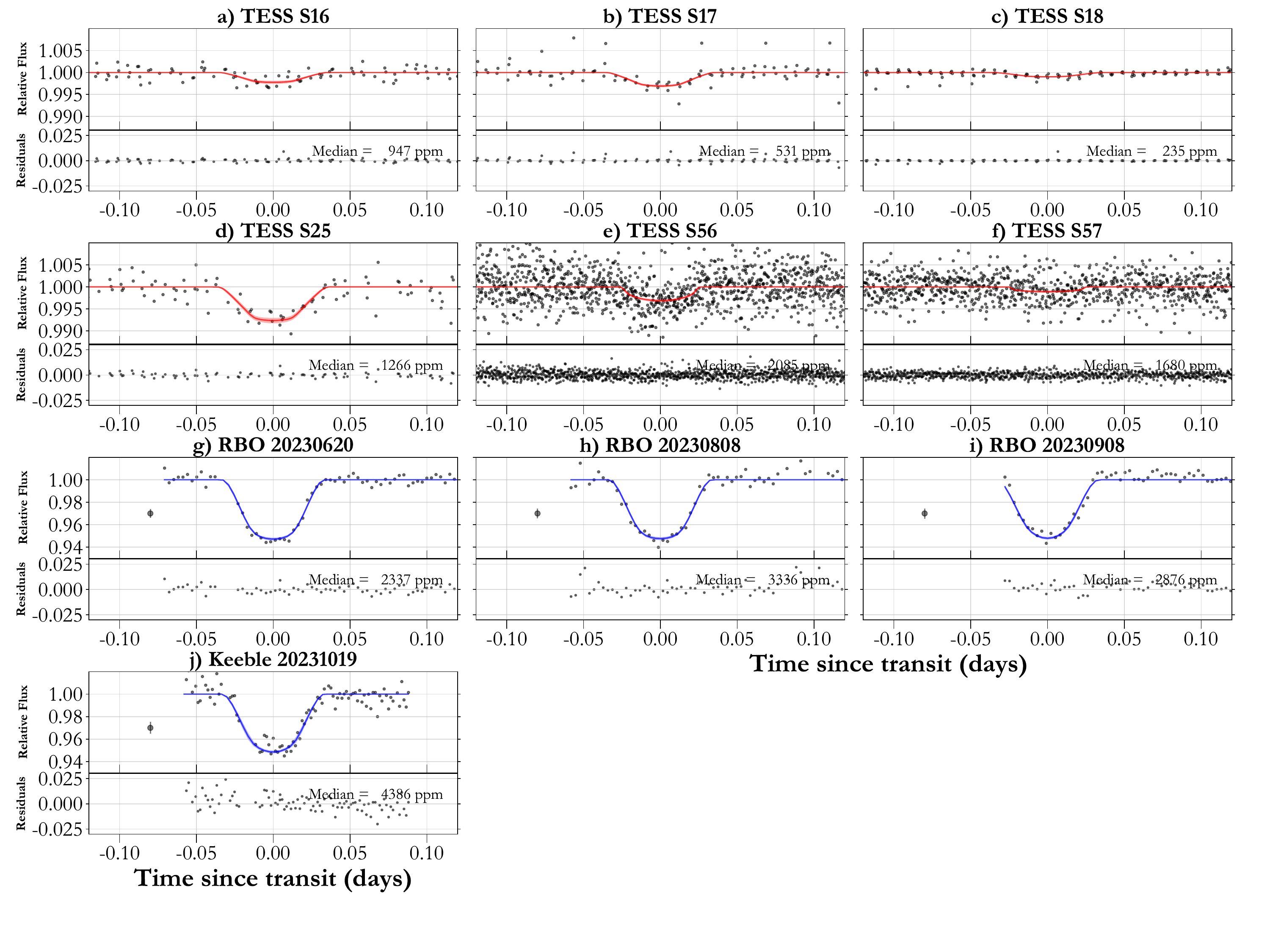}{0.95\textwidth}{}
\end{center}
\vspace{-1.5 cm}
\caption{\small \textbf{Top Left)} RV time series for TOI-6034. \textbf{Top Right)} Phase folded RVs. \textbf{Bottom)} The phase folded light curves. Note the varying TESS depths due to contamination. The plot styling and definitions are similar to \autoref{fig:Data5176}.}\label{fig:Data6034}
\end{figure*}

\begin{figure}[!t]
\begin{center}
\fig{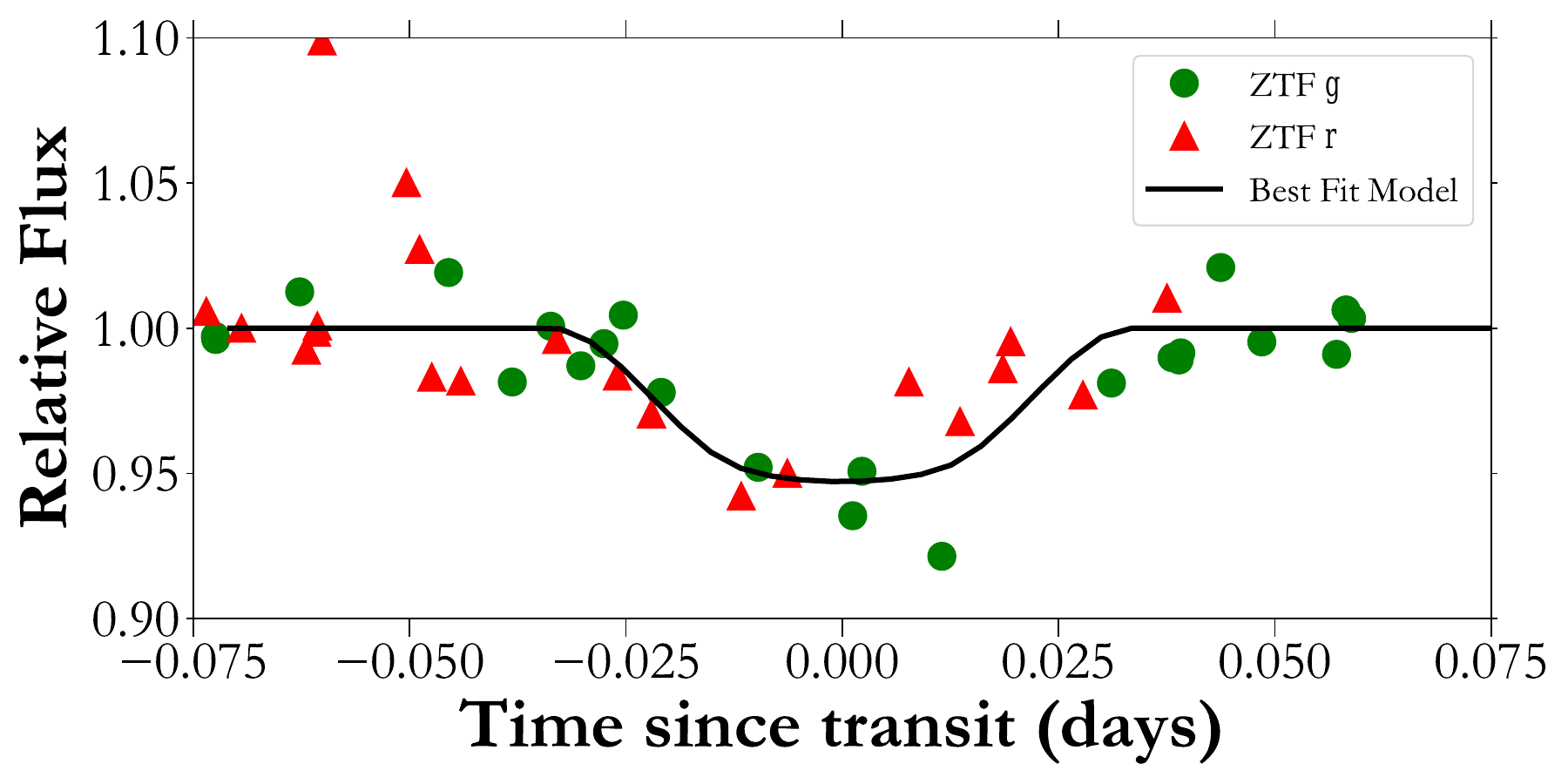}{0.95\columnwidth}{}
\vspace{-3 cm}
\fig{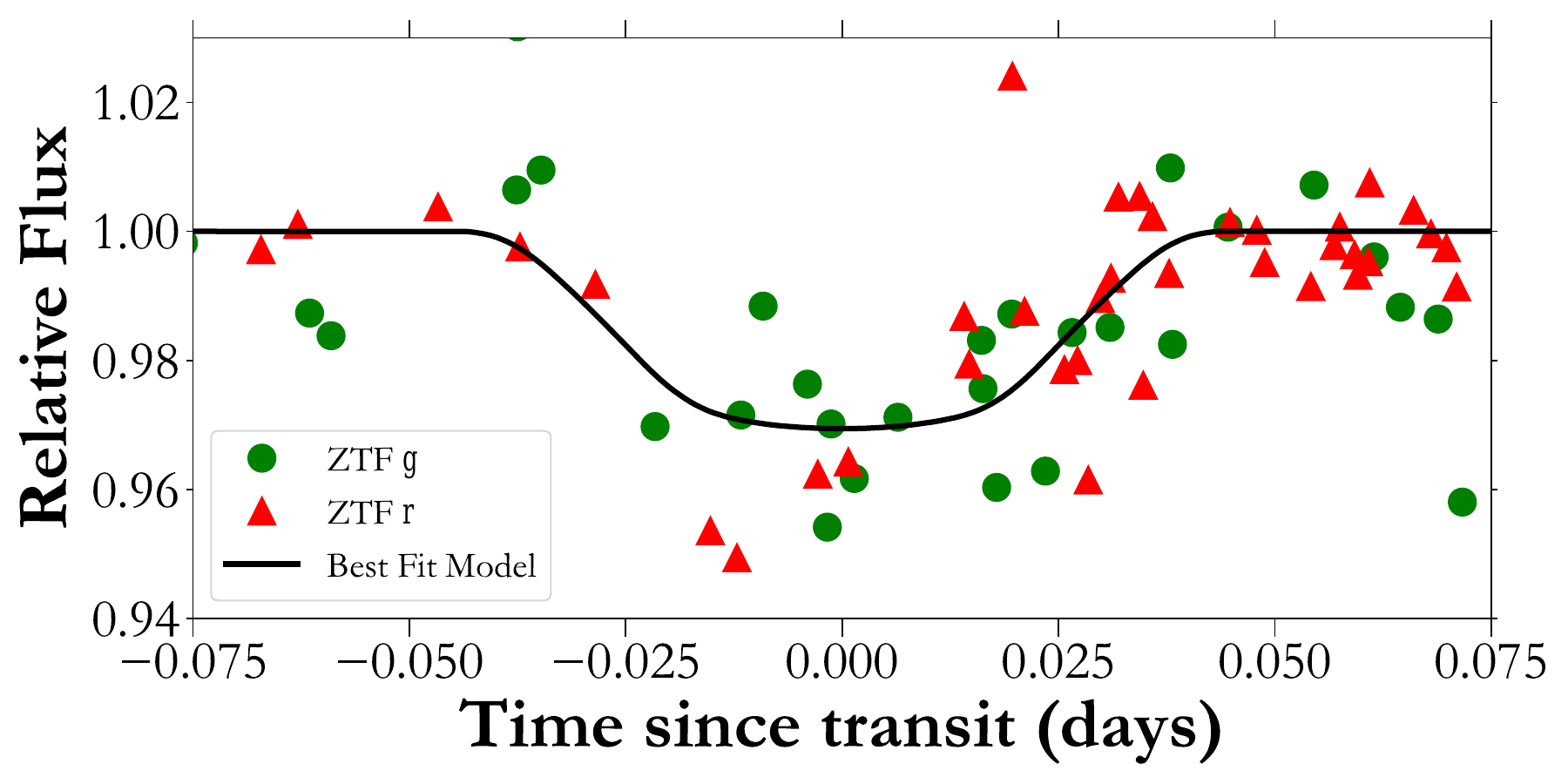}{0.95\columnwidth}{}
\end{center}
\vspace{-1 cm}
\caption{\small ZTF photometry phase folded to the best-fit planetary ephemeris, where the model is shown with a black line, and the ZTF observations in red triangles (ZTF \textit{r}) and green circles (ZTF \textit{g}). \textbf{Top:} Observations for TOI-6034 \textbf{Bottom:} Observations for TOI-5616.}\label{fig:ZTF}
\end{figure}

\textbf{TESS / \texttt{eleanor} :} TOI-6034 (TIC 388076422, 2MASS J21113603+6824074) was observed by TESS in eight sectors, which are listed in \autoref{tab:photometric}. We extract the full frame image (FFI) datasets using \texttt{eleanor} \citep{feinstein_eleanor_2019}, which uses the TESScut\footnote{\url{https://mast.stsci.edu/tesscut/}} service to obtain a cut-out of \(31\times31\) pixels. We use the \texttt{CORR\_FLUX} values for the light curve which removes correlations  with respect to the pixel position, background, and the `normal' aperture mode which uses a $3\times3$ pixel square aperture. Additionally, during the joint analysis, we use a Gaussian process to detrend the light curve (described in Section \ref{sec:joint}). 

We note that TOI-6034 has a comoving companion (TIC-388076435, Gaia DR3 2270407952771897472) about 40\arcsec\ away, which causes significant contamination in the TESS light curve and is discussed in Section \ref{sec:stellarcompanions}. This is evident in \autoref{fig:Data6034}, where we show the detrended phase-folded photometry for all the TESS sectors, with varying depth due to the large flux contamination. We discuss the procedure to correct for dilution in Section \ref{sec:joint}.  Of the eight TESS sectors, we exclude Sectors 24, 52 and 58 from analysis due to large systematic errors and contamination which preclude detection of the planet transits.

\textbf{0.6 m RBO \& 0.4~m Keeble : } We obtained three transits for TOI-6034b with the 0.6~m RBO and one with the 0.4~m Keeble observatory, which are listed in \autoref{tab:photometric} and shown in \autoref{fig:Data6034}. For RBO and Keeble, the calibration, reduction and relative photometry follows the same methodology as described in the preceding section for K2-419A.

\textbf{Zwicky Transient Facility (ZTF) / \texttt{DEATHSTAR} : } ZTF operates at the 1.2~m telescope at Palomar Observatory, California, and observers the entire Northern sky roughly every 2 days to search for transient events \citep{masci_zwicky_2019}. We use the \texttt{DEATHSTAR} \citep{ross_deathstar_2023} pipeline to extract 405 visits in ZTF \textit{g} and 525 in ZTF \textit{r} across a four and five year baseline respectively, with the parameters listed in \autoref{tab:photometric}. \texttt{DEATHSTAR} takes in the TIC ID of the star, downloads the ZTF data, extracts the lightcurves of each star in the field from the ZTF images through a custom 2-D gaussian fit, and plots the lightcurves for quick manual verification. Given the cadence, the data is not utilized during the joint fitting (Section \ref{sec:joint}), but as a qualitative check to validate the host star, and is shown in \autoref{fig:ZTF}.

\subsubsection{Radial Velocity}
We obtained 11 visits for TOI-6034 with HPF with the same exposure time and analysis protocol as detailed in Section \ref{sec:5176RVs} for K2-419A. The HPF spectra has a median S/N per pixel per unbinned exposure (of 969 s) at 1070 nm of 46. The HPF RVs for TOI-6034~A are listed in \autoref{tab:rvs}.

\subsubsection{High-contrast Imaging}

\begin{figure*}[!t]
\begin{center}
\fig{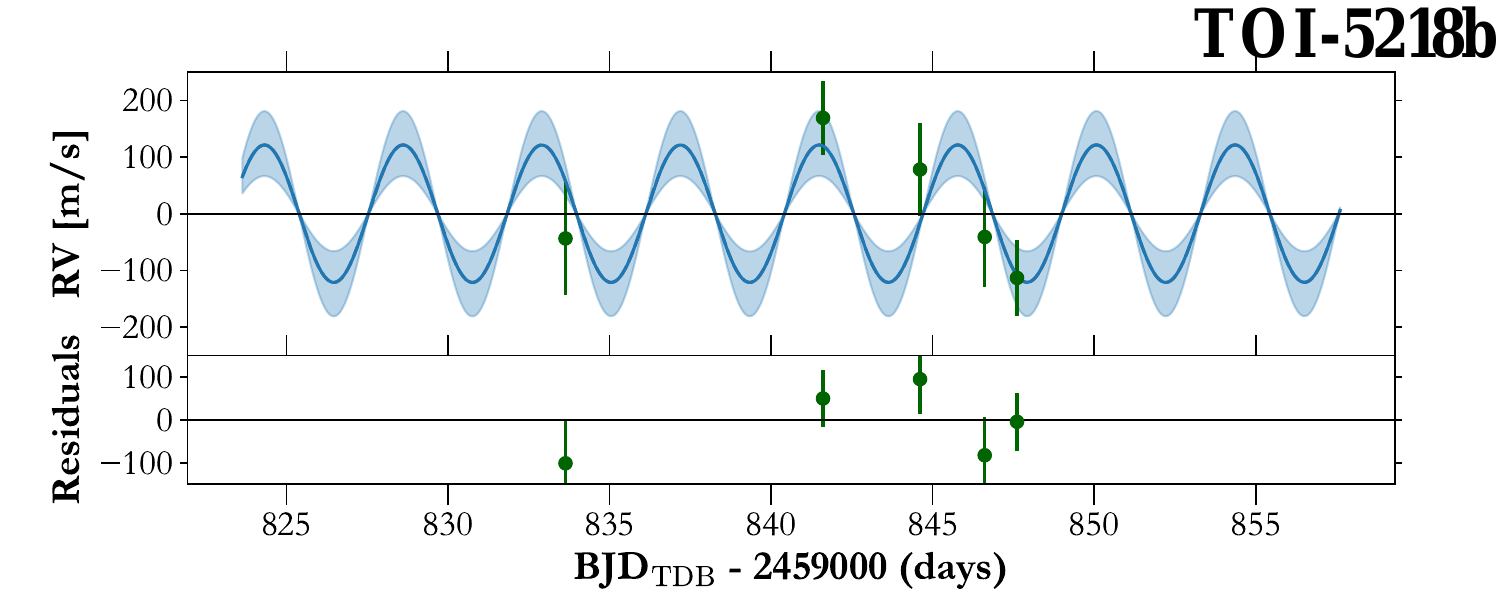}{0.62\textwidth}{}
\fig{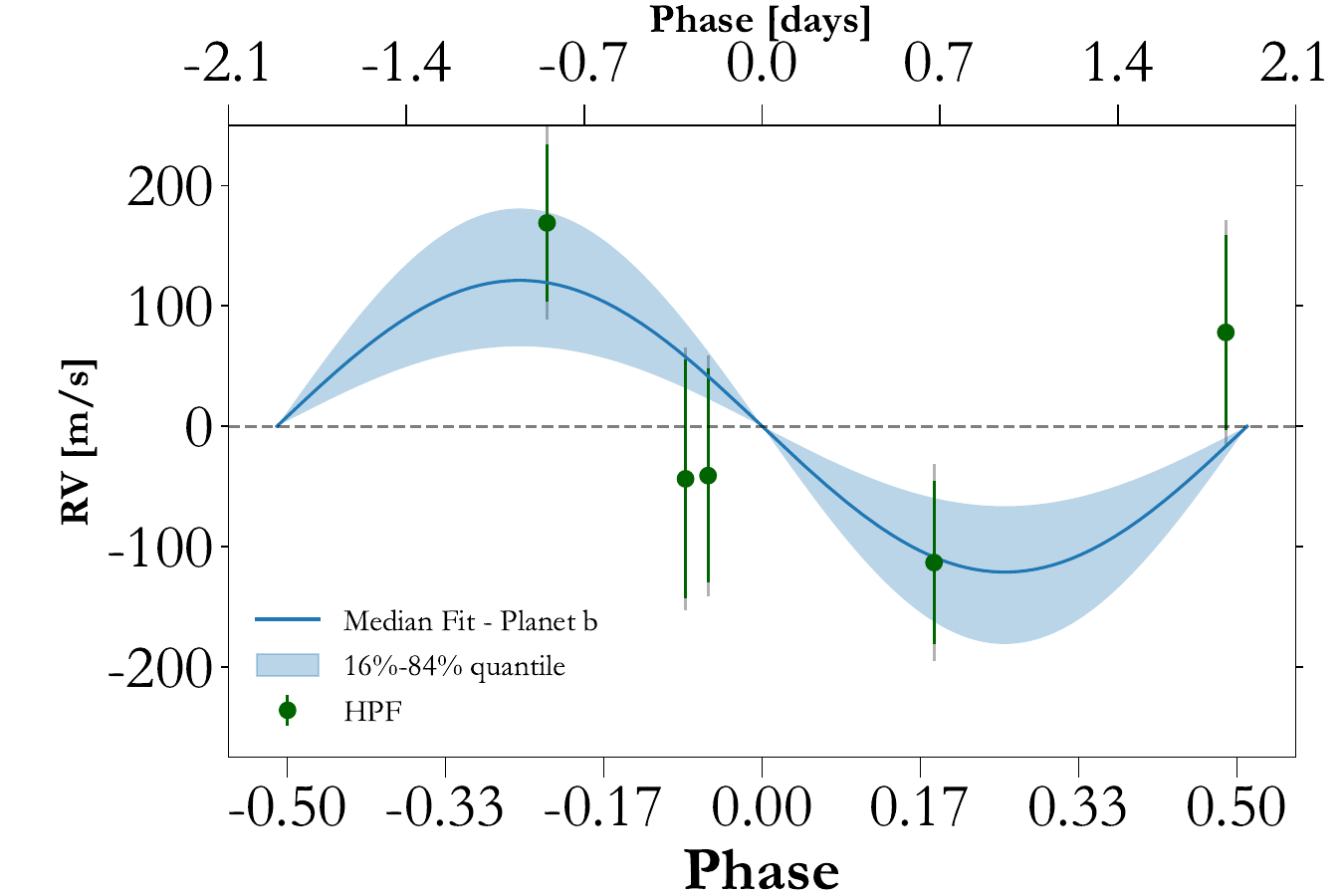}{0.36\textwidth}{}
\fig{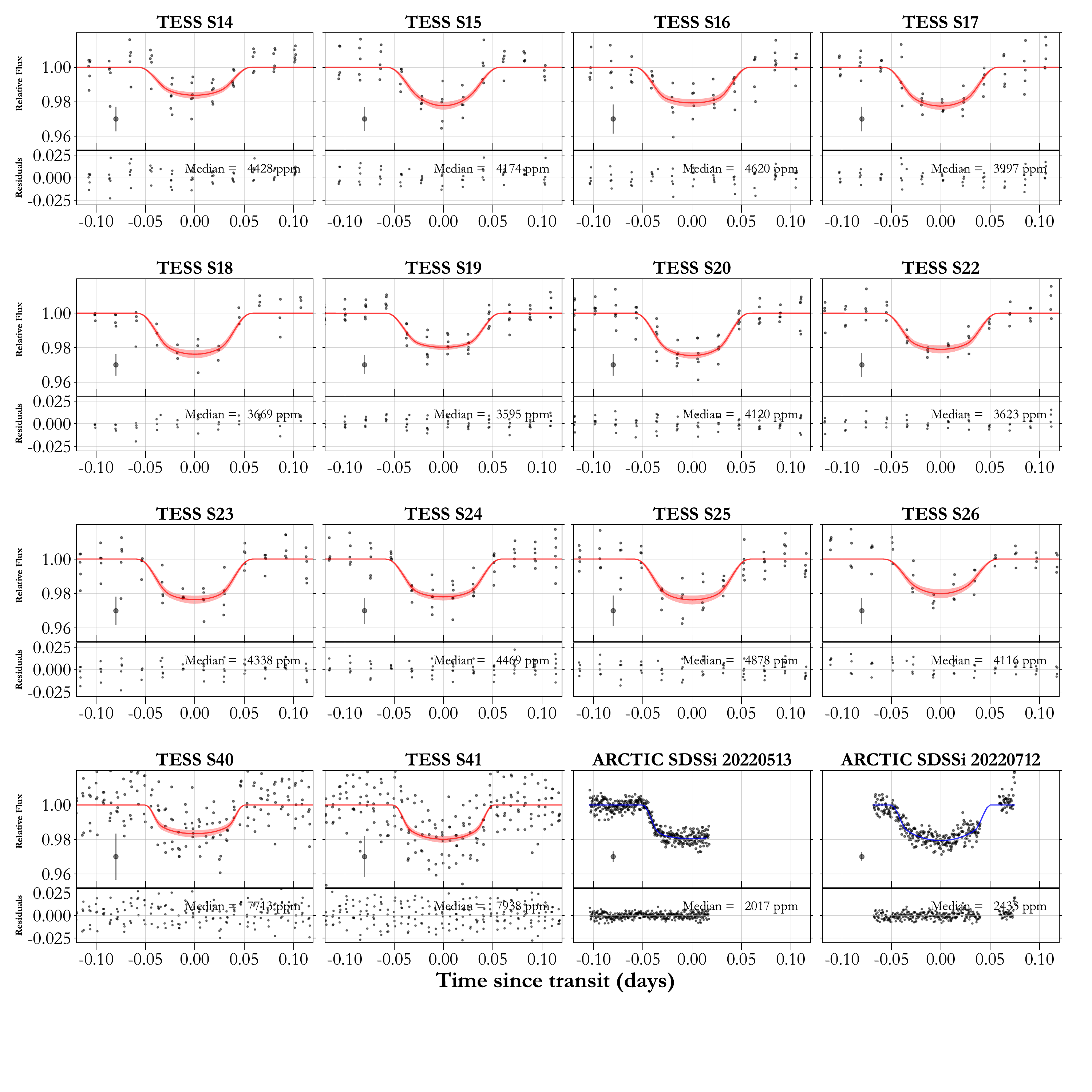}{0.95\textwidth}{}
\end{center}
\vspace{-2 cm}
\caption{\small \textbf{Top Left)} RV time series for TOI-5218. \textbf{Top Right)} Phase folded RVs. \textbf{Bottom)} The phase folded light curves. The plot  and definitions are similar to \autoref{fig:Data5176}.}\label{fig:Data5218}
\end{figure*}

TOI-6034 was observed with NESSI on 2023 July 18. Speckle data were taken simultaneously using the red and blue NESSI cameras in the SDSS \textit{z}' and \textit{r}' filters, respectively. Images were reconstructed as described in Section \ref{sec:5176HC}. The $5\sigma$ contrast limits and the z' image are shown in \autoref{fig:HighContrast}. These data rule out the presence of any nearby sources brighter than $\Delta z' = 2.7$ mag or $\Delta r' = 3.1$ mag at 0.2\arcsec\ and $\Delta z' = 4.2$ mag or $\Delta r' = 3.3$ mag at 1.2\arcsec.

Analogous speckle data were also taken with the `Alopeke  imager \citep{scott_twin_2021} on the Gemini North 8.1~m on Mauna Kea, using the 832 nm and 562 nm filters on the red and blue cameras. The observations were carried out on 2022 September 13,  and data processing followed the same procedures described for NESSI, with which `Alopeke shares a similar design. The $5\sigma$ contrast limits are shown in \autoref{fig:HighContrast}. The `Alopeke data rule out nearby sources to a limit of $\Delta {\rm m}_{832} = 4.9$ mag or $\Delta {\rm m}_{562} = 4.4$ mag at 0.2\arcsec\ and $\Delta {\rm m}_{832} = 5.5$ mag or $\Delta {\rm m}_{562} = 4.7$ mag at 1.15\arcsec.

We also observed TOI-6034 with the ShARCS camera on the Shane 3~m telescope at Lick Observatory \citep{srinath_swimming_2014}. The target was observed on 2024 May 26 with fluctuating seeing conditions between 0.9\arcsec and 1.6\arcsec over the night. It was observed in Laser Guide Star (LGS) mode at a single position for a total of 1500s exposure time and then manually reduced with offset sky frames taken between exposures. We use the algorithm developed by \cite{espinoza_hats-25b_2016} to generate a 5 sigma contrast curve as a part of the final analysis (\autoref{fig:HighContrast}). We detected no companions outside $>$ 0.99\arcsec corresponding to a $\Delta$Ks of 2.78, or within $>$ 1.5\arcsec corresponding to a $\Delta$Ks of 4.44.

\subsection{TOI-5218}
\subsubsection{Photometry}\label{sec:5218phot}
\textbf{TESS / \texttt{tglc} :} TOI-5218 (TIC-259172249, 2MASS J19303919+7145463) was observed by TESS in 26 sectors up until Year 5 of TESS' observing, which are listed in \autoref{tab:photometric}. Given the need for detrending, and computational tractability, we use the first 15 sectors of TESS observations for TOI-5218.  We use the TESS-Gaia Light Curve \citep[\texttt{tglc}; ][]{han_tess-gaia_2023} package to extract TESS photometry from a $3 \times 3$ pixel aperture, after estimating the effective PSF using a $90 \times 90$ pixel FFI cutout. \texttt{tglc} models the point spreads function of all \gaia~stars using a combination of \gaia~positions, proper motions, and colour information, which is subsequently used to remove contamination from all stars other than the target. It also removes major CCD artifacts and the gradient in the field caused by stray light.

\textbf{ARC 3.5~m / ARCTIC :} We observed TOI-5218 on the nights of 2022 May 13 and 2022 July 12 in SDSS~\textit{i'} using the $3.5\unit{m}$ Astrophysical Research Consortium (ARC) Telescope Imaging Camera \citep[ARCTIC;][]{huehnerhoff_astrophysical_2016} at the ARC 3.5m Telescope at Apache Point Observatory (APO). Given the presence of a background companion $\sim$ 5\arcsec~away, we defocus moderately to 2.8 \arcsec{} on both nights. The observations were conducted in the $4 \times 4$ binning mode, with a gain of $2.0 \unit{e^-/ADU}$, a plate scale of $0.456 \unit{\arcsec/pixel}$, and a readout time of  $1.3 \unit{s}$. The data was calibrated and reduced using \texttt{AIJ}, and is listed in \autoref{tab:photometric}, and shown in \autoref{fig:Data5218}.

\subsubsection{Radial Velocity}
We obtained 5 visits for TOI-5218 with HPF with the same exposure time and analysis protocol as detailed in Section \ref{sec:5176RVs} for K2-419A. The HPF spectra has a median S/N per pixel per unbinned exposure (of 969 s) at 1070 nm of 19. The HPF RVs for TOI-5218 are listed in \autoref{tab:rvs}.

\subsubsection{High-contrast Imaging}

TOI-5218 was observed with NESSI on 2022 September 16. Speckle data were taken simultaneously using the red and blue NESSI cameras in the SDSS \textit{z}' and \textit{r}' filters, respectively. Images were reconstructed as described in Section \ref{sec:5176HC}. The $5\sigma$ contrast limits and the z' image are shown in \autoref{fig:HighContrast}. These data rule out the presence of any nearby sources brighter than $\Delta z' = 3.5$ mag or $\Delta r' = 3.3$ mag at 0.2\arcsec\ and $\Delta z' = 4.2$ mag or $\Delta r' = 4.0$ mag at 1.2\arcsec.

\subsection{TOI-5414}

\begin{figure*}[!t]
\begin{center}
\fig{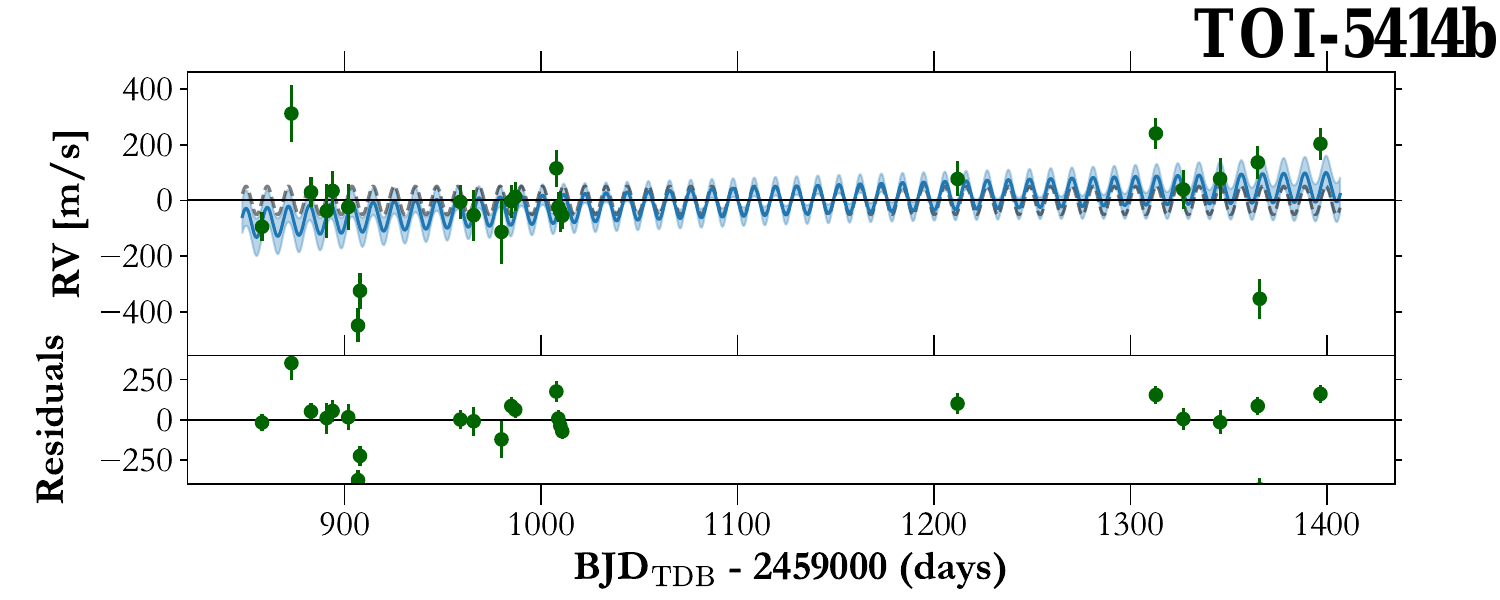}{0.62\textwidth}{}
\fig{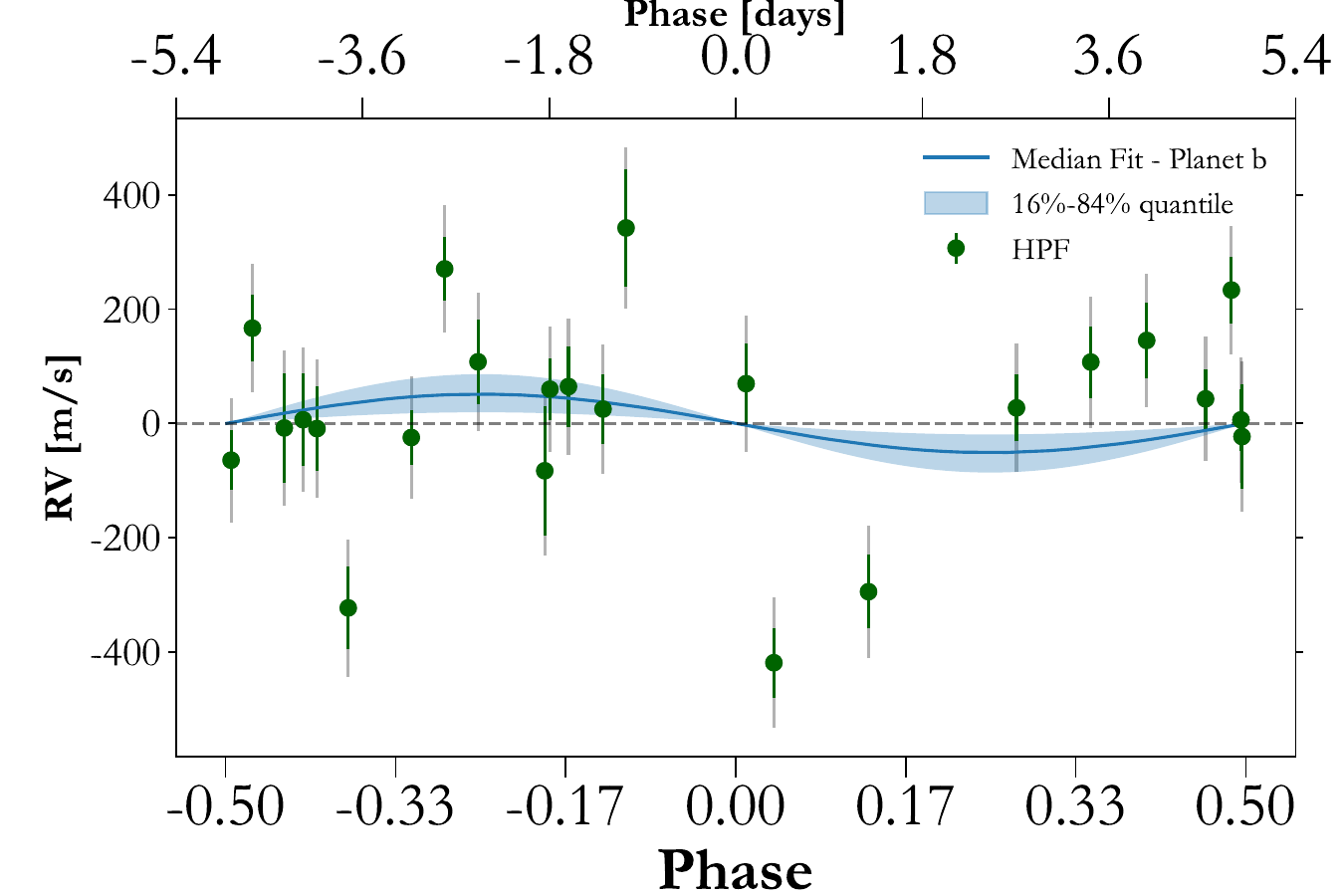}{0.36\textwidth}{}
\fig{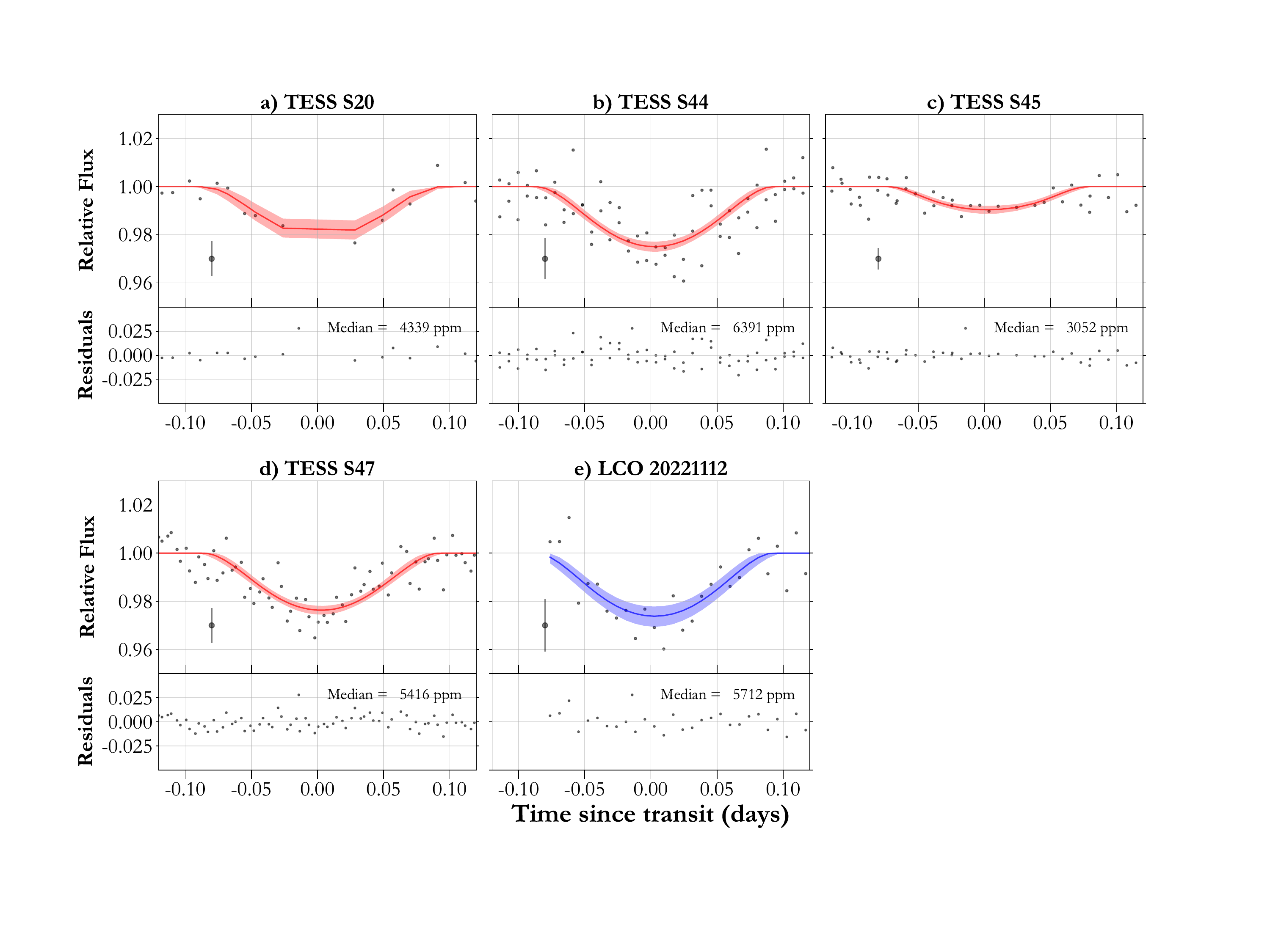}{0.95\textwidth}{}
\end{center}
\vspace{-2 cm}
\caption{\small \textbf{Top Left)} RV time series for TOI-5414. \textbf{Top Right)} Phase folded RVs. \textbf{Bottom)} The phase folded light curves. The plot styling and definitions are similar to \autoref{fig:Data5176}.}\label{fig:Data5414}
\end{figure*}

\subsubsection{Photometry}\label{sec:5414phot}
\textbf{TESS / \texttt{eleanor} :} TOI-5414 (TIC 371164829, 2MASS J07172825+3128550) was observed by TESS in four sectors, which are listed in \autoref{tab:photometric}. We extract the the photometry using \texttt{eleanor} similar to that for TOI-6034 in Section \ref{sec:6034phot}.

\textbf{0.4 m LCOGT Archival Data :}
We utilize a transit of TOI-5414b from the Las Cumbres Observatory global telescope network \citep[LCOGT;][]{brown_cumbres_2013} archive, as observed on 2022 November 12 as part of proposal KEY2020B-005 (PI: Shporer, A.) from the 0.4 m Planewave Delta Rho telescope with a QHY 600 CMOS camera at the Teide Observatory in Tenerife, Spain. The observation parameters are described in \autoref{tab:photometric}, with the raw data processed by the BANZAI pipeline \citep{mccully_real-time_2018}, and processed through \texttt{AIJ}.

\subsubsection{Radial Velocity}
We obtained 24 visits for TOI-5414 with HPF with the same exposure time and analysis protocol as detailed in Section \ref{sec:5176RVs} for K2-419A. The HPF spectra has a median S/N per pixel per unbinned exposure (of 969 s) at 1070 nm of 22. The HPF RVs for TOI-5414 are listed in \autoref{tab:rvs}.

\subsubsection{High-contrast Imaging}\label{sec:5414HC}

TOI-5414 was observed with NESSI on 2024 February 15. Speckle data were taken simultaneously using the red and blue NESSI cameras in the SDSS \textit{z}' and \textit{r}' filters, respectively. Images were reconstructed as described in Section \ref{sec:5176HC}. The $5\sigma$ contrast limits and the z' image are shown in \autoref{fig:HighContrast}. These data rule out the presence of any nearby sources brighter than $\Delta z' = 3.4$ mag or $\Delta r' = 3.9$ mag at 0.2\arcsec\ and $\Delta z' = 4.2$ mag or $\Delta r' = 4.6$ mag at 1.2\arcsec.

Observations of TOI-5414 were made on 2023 November 26 with the PHARO instrument \citep{hayward2001} on the Palomar Hale (5m) behind the P3K natural guide star AO system \citep{dekany_palm-3000_2013} in the narrowband K$_{cont}$ filter $(\lambda_o = 2.29; \Delta\lambda = 0.035~\mu$m). The PHARO pixel scale is $0.025\arcsec$ per pixel. A standard 5-point dither pattern with steps of 5\arcsec\ was repeated twice with each repeat separated by 0.5\arcsec.  Flat fields were taken on-sky, dark-subtracted, and median averaged, and sky frames were generated from the median average of the dithered science frames. Each science image was then sky-subtracted and flat-fielded.  The reduced science frames were combined into a single mosaic-ed image.  The final resolution of the combined dither was determined from the full-width half-maximum (FWHM) of the point spread function (PSF): 0.16\arcsec.
	
The sensitivities of the final combined AO image were determined by injecting simulated sources azimuthally around the primary target every $20^\circ $ at separations of integer multiples of the central source's FWHM \citep{furlan_kepler_2017}. The brightness of each injected source was scaled until standard aperture photometry detected it with $5\sigma $ significance.  The final $5\sigma $ limit at each separation was determined from the average of all of the determined limits at that separation and the uncertainty on the limit was set by the rms dispersion of the azimuthal slices at a given radial distance.  The sensitivity limits and PHARO image cutout are shown in \autoref{fig:HighContrast}.


\subsection{TOI-5616}

\begin{figure*}[!t]
\begin{center}
\fig{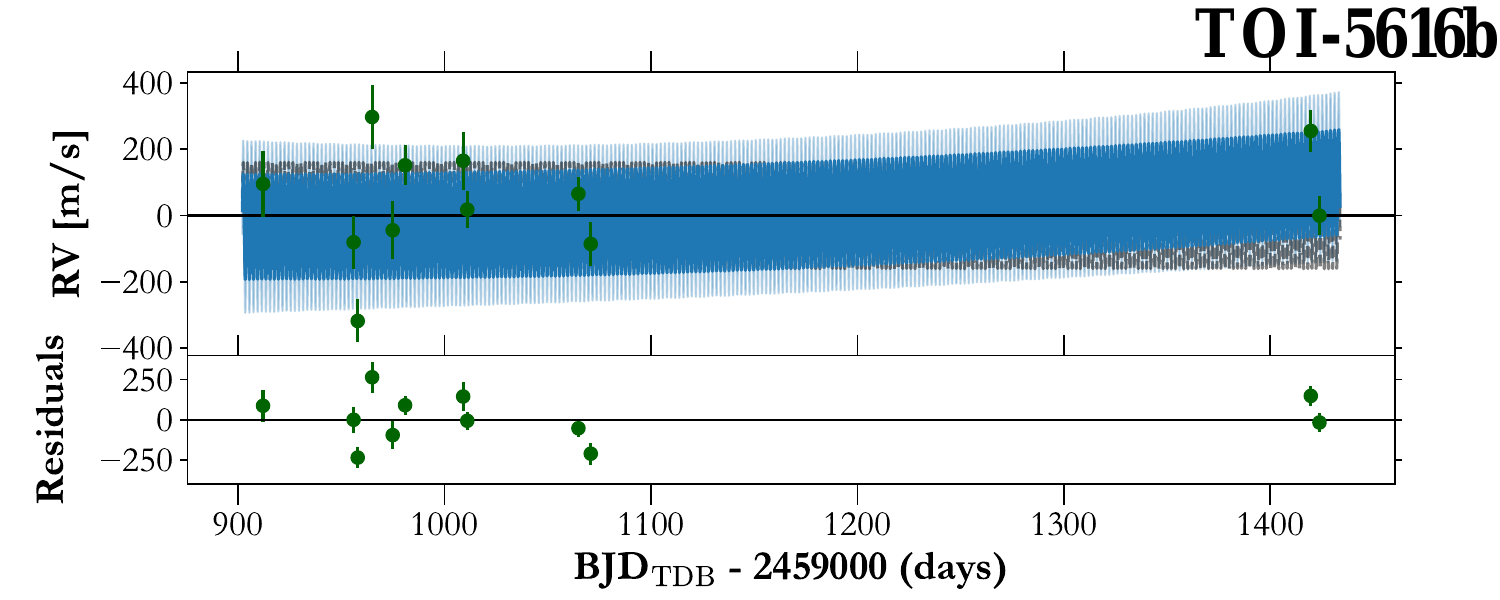}{0.62\textwidth}{}
\fig{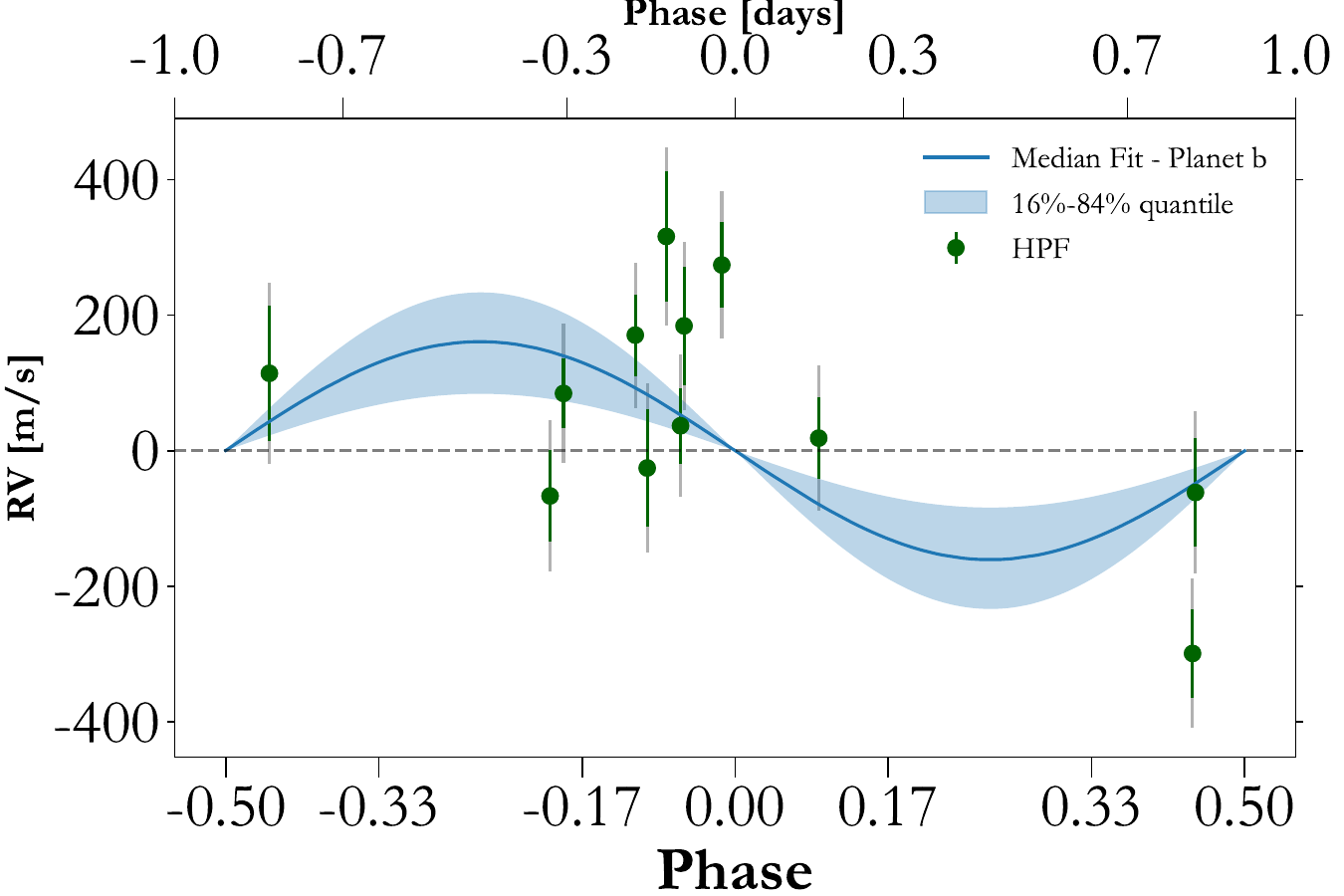}{0.36\textwidth}{}
\fig{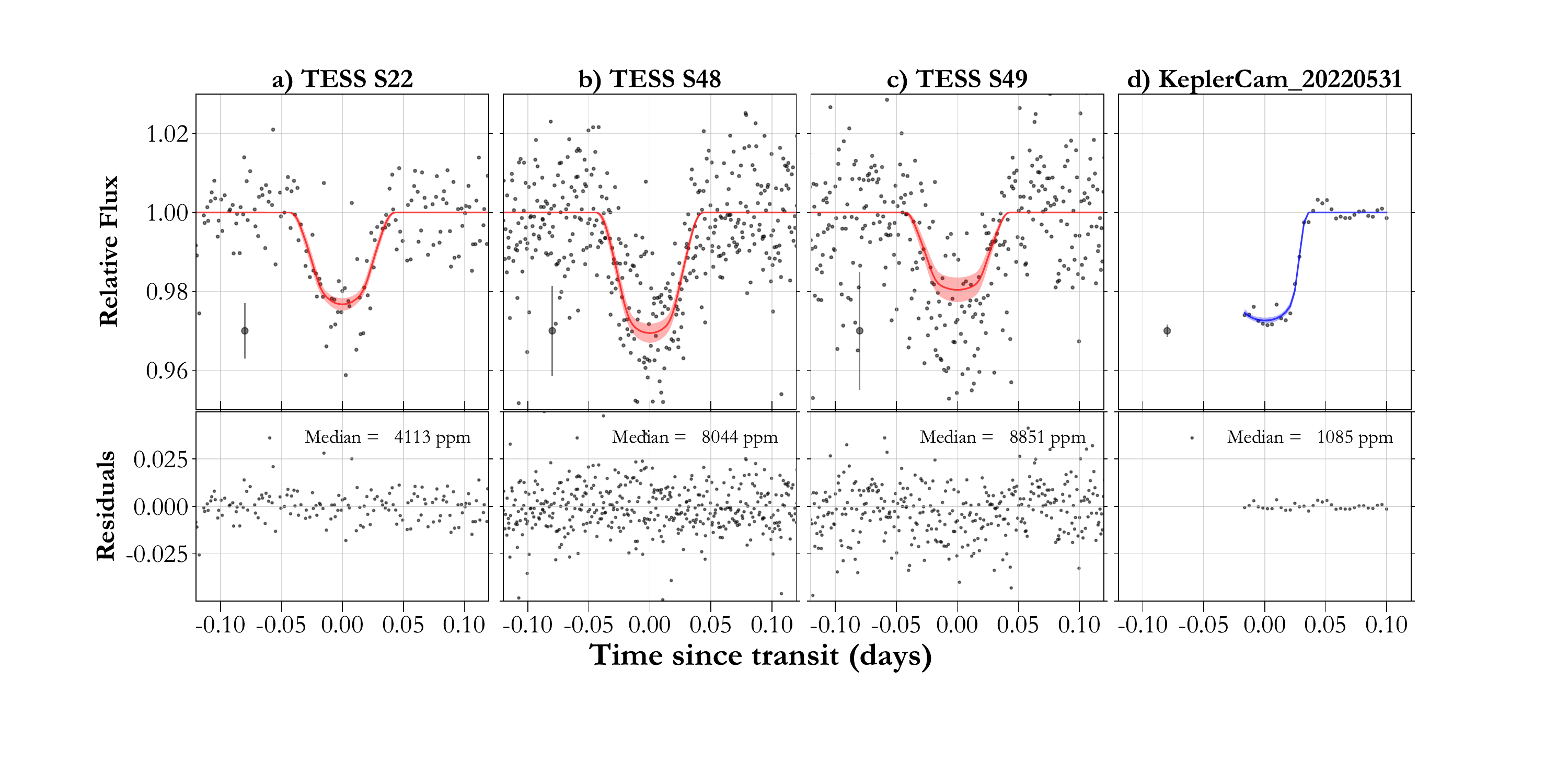}{0.95\textwidth}{}
\end{center}
\vspace{-2 cm}
\caption{\small \textbf{Top Left)} RV time series for TOI-5616. \textbf{Top Right)} Phase folded RVs. \textbf{Bottom)} The phase folded light curves. The plot styling and definitions are similar to \autoref{fig:Data5176}.}\label{fig:Data5616}
\end{figure*}

\subsubsection{Photometry}\label{sec:5616phot}
\textbf{TESS / \texttt{tglc} :} TOI-5616 (TIC-154220877, 2MASS J12193646+4840593) was observed by TESS in three sectors, which are listed in \autoref{tab:photometric}. We use \texttt{tglc} to extract the light curve from the TESS FFIs similar to that described for TOI-5218 in Section \ref{sec:5218phot}, while using a 31 $\times$ 31 pixel FFI cutout.

\textbf{1.2 m FLWO / KeplerCam :} We obtained a transit of TOI-5616 from KeplerCam on the 1.2 m \textit{Fred Lawrence Whipple Observatory} (FLWO) in southern Arizona, USA \citep{szentgyorgyi_keplercam_2005} on 2022 May 31 (\autoref{tab:photometric}). It has a 4096$\times$4096 FairChild CCD with a 0.67\arcsec~plate-scale per 2$\times$2 binned pixel. The data was calibrated and reduced using \texttt{AIJ}, and is shown in \autoref{fig:Data5616}.

\textbf{Zwicky Transient Facility (ZTF) / \texttt{DEATHSTAR} : } Similar to TOI-6034, we also utilize 529 visits in ZTF \textit{g} and 568 in ZTF \textit{r} that are listed in \autoref{tab:photometric}, and shown in \autoref{fig:ZTF}.  Given the cadence, the data is not utilized during the joint fitting (Section \ref{sec:joint}), but as a qualitative check to validate the host star, and is shown in \autoref{fig:ZTF}.

\subsubsection{Radial Velocity}
We obtained 12 visits for TOI-5616 with HPF with the same exposure time and analysis protocol as detailed in Section \ref{sec:5176RVs} for K2-419A. The HPF spectra has a median S/N per pixel per unbinned exposure (of 969 s) at 1070 nm of 19. The HPF RVs for TOI-5616 are listed in \autoref{tab:rvs}.

\subsubsection{High-contrast Imaging}

TOI-5616 was observed with NESSI on 2023 January 28. Speckle data were taken simultaneously using the red and blue NESSI cameras in the SDSS \textit{z'} and \textit{r'} filters, respectively. Images were reconstructed as described in Section \ref{sec:5176HC}. The $5\sigma$ contrast limits and the z' image are shown in \autoref{fig:HighContrast}. These data rule out the presence of any nearby sources brighter than $\Delta z' = 3.6$ mag or $\Delta r' = 2.9$ mag at 0.2\arcsec\ and $\Delta z' = 4.7$ mag or $\Delta r' = 4.2$ mag at 1.2\arcsec.


TOI-5616 was also observed with PHARO on 2024 April 22. The data were processed following the same methods described in Section \ref{sec:5414HC} for TOI-5414, though with a final resolution corresponding to a PSF FWHM of 0.11\arcsec. The sensitivity limits and PHARO image cutout are shown in \autoref{fig:HighContrast}.

\subsection{TOI-5634A}

\begin{figure*}[!t]
\begin{center}
\fig{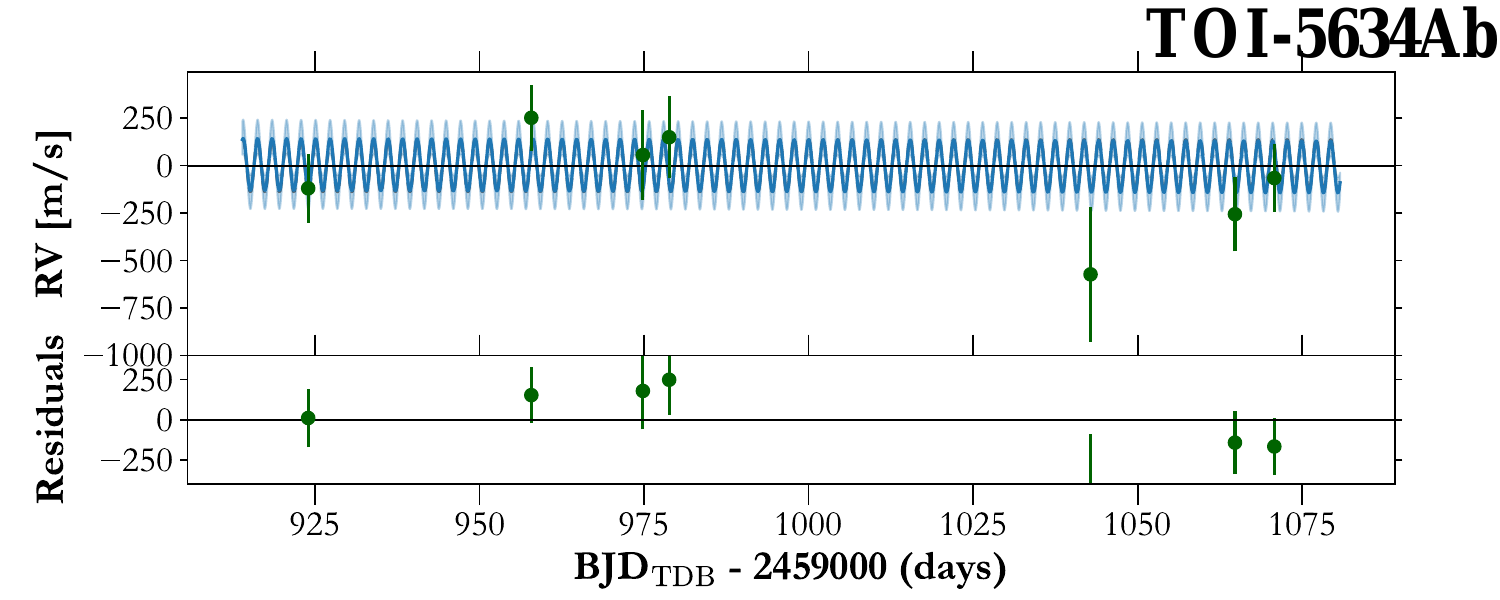}{0.62\textwidth}{}
\fig{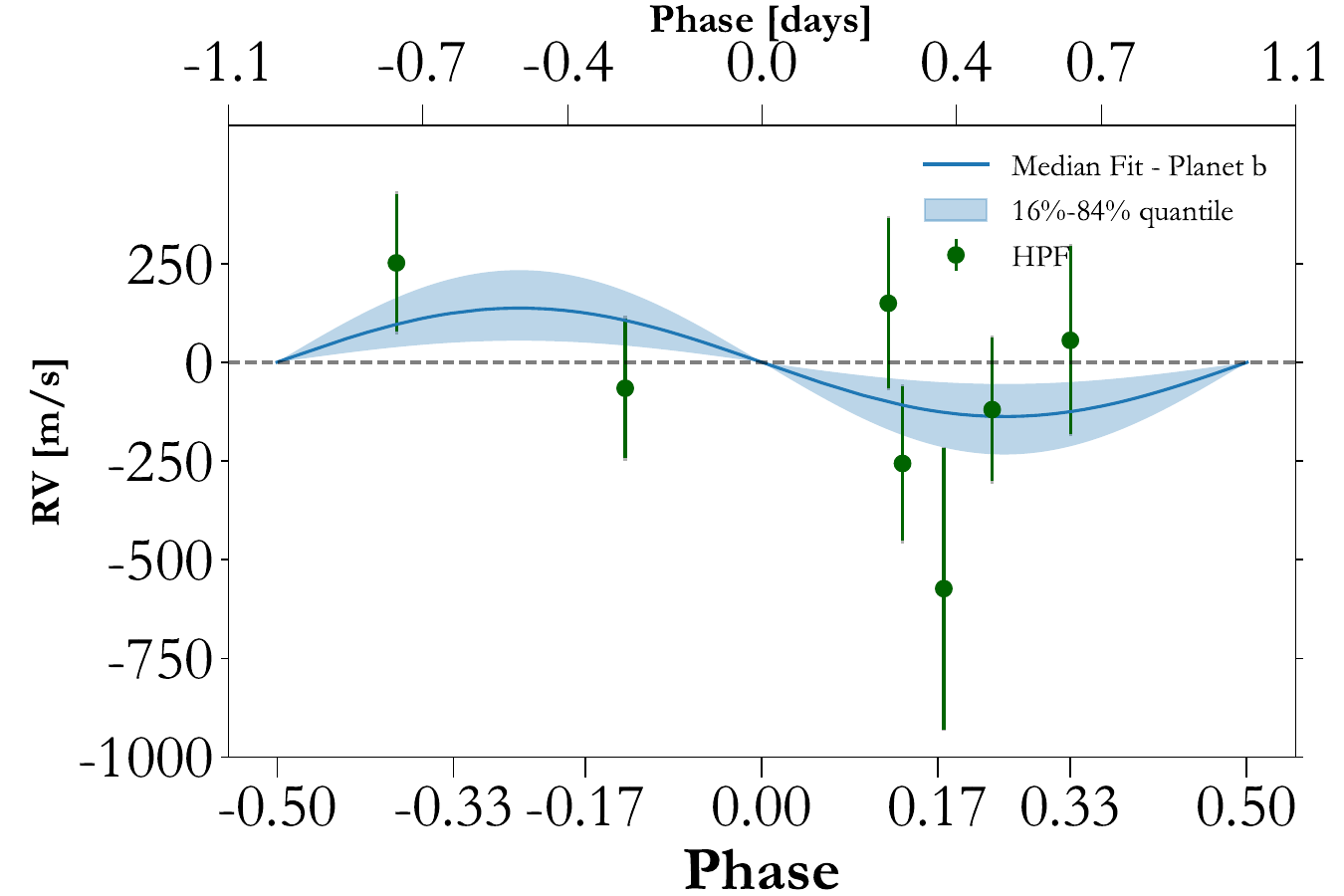}{0.36\textwidth}{}
\fig{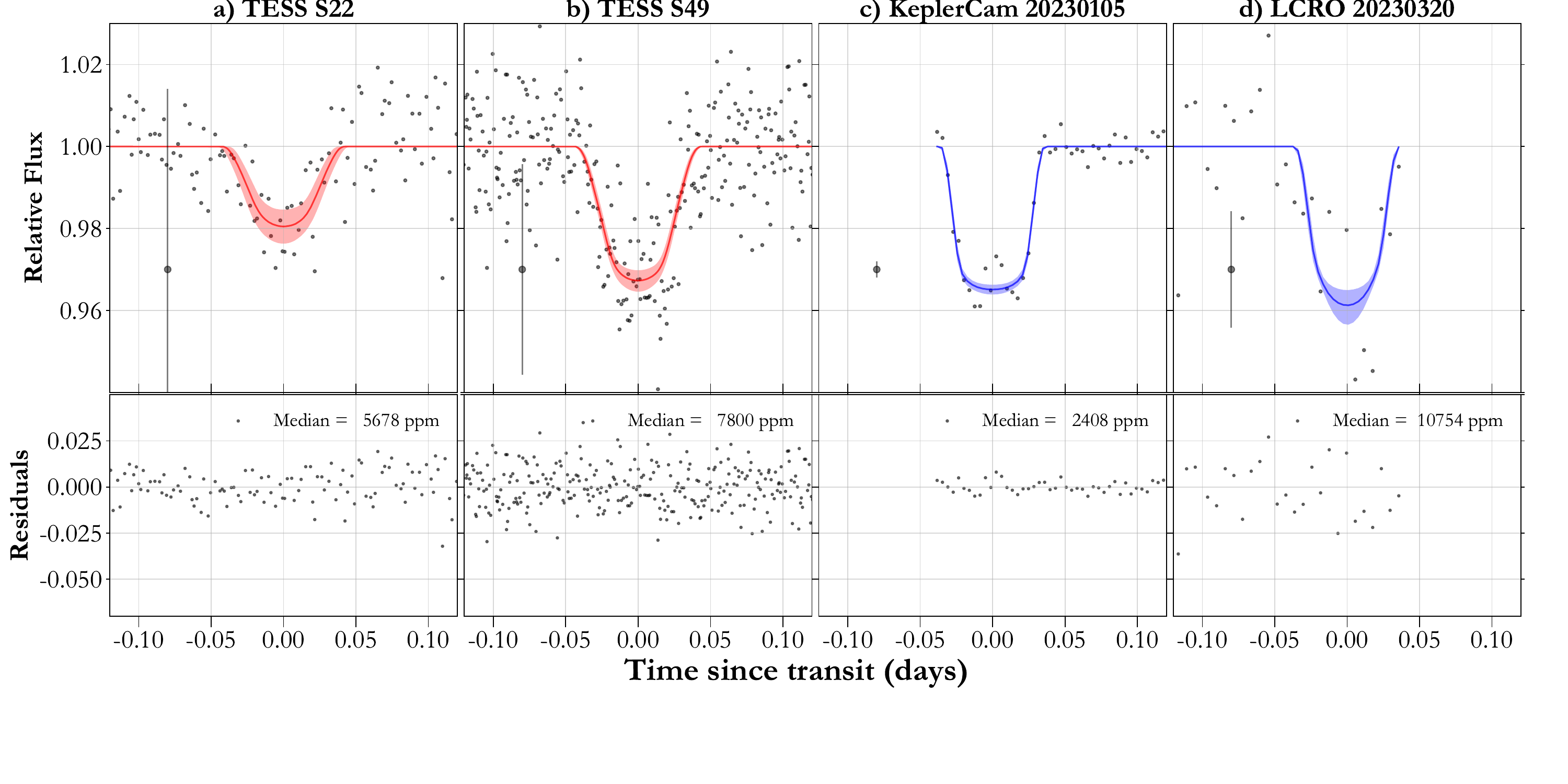}{0.95\textwidth}{}
\end{center}
\vspace{-2 cm}
\caption{\small \textbf{Top Left)} RV time series for TOI-5634A. \textbf{Top Right)} Phase folded RVs. \textbf{Bottom)} The phase folded light curves. The plot styling and definitions are similar to \autoref{fig:Data5176}.}\label{fig:Data5634}
\end{figure*}

\subsubsection{Photometry}\label{sec:5634phot}
\textbf{TESS / \texttt{tglc} :} TOI-5634A (TIC-119585136, 2MASS J11423454+2057553) was observed by TESS in two sectors that are listed in \autoref{tab:photometric}. We use the \texttt{tglc} package to extract the photometry similar to that for TOI-5616. The extracted, detrended and phase folded TESS photometry is shown in \autoref{fig:Data5634}.

\textbf{1.2 m FLWO / KeplerCam :} We obtained a transit of TOI-5634Ab on 2023 January 5, the details of which are included in \autoref{tab:photometric}. The data was processed using AIJ, similar to that for TOI-5616 described in Section \ref{sec:5616phot}.

\textbf{0.3 m LCRO :} We obtained a transit for TOI-5634Ab on 2023 March 20 using the $305 \unit{mm}$ Las Campanas Remote Observatory (LCRO) telescope at the Las Campanas Observatory in Chile. This is an f/8 Maksutov-Cassegrain from Astro-Physics on a German Equatorial AP1600 GTO mount with an FLI Proline 16803 CCD Camera, FLI ATLAS focuser and Centerline filter wheel. The data was taken unbinned, with a plate scale of 0.773\arcsec per pixel. The data was analyzed using the same custom \texttt{python} pipeline used with the 0.6~m RBO and described in Section \ref{sec:5176phot}.

\subsubsection{Radial Velocity}
We obtained 7 visits for TOI-5634A with HPF with the same exposure time and analysis protocol as detailed in Section \ref{sec:5176RVs} for K2-419A. The HPF spectra has a median S/N per pixel per unbinned exposure (of 969 s) at 1070 nm of 16. The HPF RVs for TOI-5634A are listed in \autoref{tab:rvs}.

\subsubsection{High-contrast Imaging}

TOI-5634A was observed with NESSI on 2023 January 28. Speckle data were taken simultaneously using the red and blue NESSI cameras in the SDSS \textit{z'} and \textit{r'} filters, respectively. Images were reconstructed as described in Section \ref{sec:5176HC}. The $5\sigma$ contrast limits and the z' image are shown in \autoref{fig:HighContrast}. These data rule out the presence of any nearby sources brighter than $\Delta z' = 3.7$ mag or $\Delta r' = 4.3$ mag at 0.2\arcsec\ and $\Delta z' = 4.6$ mag or $\Delta r' = 4.8$ mag at 1.2\arcsec.


TOI-5634A was also observed with PHARO on 2024 February 15. The data were processed following the same methods described in \ref{sec:5414HC} for TOI-5414, though with a final resolution corresponding to a PSF FWHM of 0.24\arcsec. The sensitivity limits and PHARO image cutout are shown in \autoref{fig:HighContrast}.
This star has a detected companion 4\arcsec\ to the southwest that is likely to be bound.  The source (TIC 903545876 = Gaia DR3 3979511431397963008) is also detected by Gaia DR2 and DR3 and has the approximately the same distance and proper motion ($d\approx334\pm30$ pc; pmra $\approx-60.8\pm0.2$ mas~yr$^{-1}$; pmdec $\approx-25.6\pm0.2$ mas~yr$^{-1}$) as  TOI-5634A. The companion star is an early/mid M-dwarf (M2.5V) with a temperature of $\sim 3400$ K; at a separation of 4\arcsec\, the companion star is approximately 1300 au in projected separation. These two stars are also deemed to be a bound pair based on their \textit{Gaia} astrometry by \cite{el-badry_million_2021}. Based on this we refer to the planet host as TOI-5634A.

\begin{deluxetable*}{lccccc}
\tabletypesize{\fontsize{8}{11}\selectfont}
\tablecaption{Summary of stellar parameters for the TOIs. \label{tab:stellarparam}}
\tablehead{
\colhead{~~~Parameter}&  
\colhead{Description}&
\colhead{K2-419A}&
\colhead{TOI-6034}&
\colhead{TOI-5218}&
\colhead{Reference}
}
\startdata
\multicolumn{4}{l}{\hspace{-0.2cm} Main identifiers:}  \\
~~~TOI & \tess{} Object of Interest & 5176 & 6034 & 5218 & \tess{} mission \\
~~~TIC & \tess{} Input Catalogue  & 437054764 & 388076422 & 259172249 &  Stassun \\
~~~2MASS & \(\cdots\) & J09000476+1316258 &  J21113603+6824074 &  J19303919+7145463  & 2MASS  \\
~~~Gaia DR3 & \(\cdots\) & 605593554127479936 & 2270404997834401664 & 2263568371970517120 & Gaia DR3\\
\multicolumn{4}{l}{\hspace{-0.2cm} Equatorial Coordinates, Proper Motion and Spectral Type:} \\
~~~$\alpha_{\mathrm{J2016}}$ &  Right Ascension (RA) & 135.01948 $\pm$ 0.0317 & 317.9001 $\pm$ 0.0149 & 292.6635 $\pm$ 0.0232 & Gaia DR3\\
~~~$\delta_{\mathrm{J2016}}$ &  Declination (Dec) & 13.2737 $\pm$ 0.0232 & 68.4021 $\pm$ 0.0159 & 71.7630 $\pm$ 0.0223 & Gaia DR3\\
~~~$\mu_{\alpha}$ &  Proper motion (RA, \unit{mas/yr}) & -71.336 $\pm$ 0.042 & -2.430 $\pm$ 0.020 & 8.038 $\pm$ 0.032  & Gaia DR3\\
~~~$\mu_{\delta}$ &  Proper motion (Dec, \unit{mas/yr}) & -27.146 $\pm$ 0.033 & -5.109 $\pm$ 0.021 & 17.2503 $\pm$ 0.030 & Gaia DR3 \\
~~~$d$ &  Distance in pc  & 263.8 $\pm$ 2.5 & 117.48 $\pm$ 0.22 & 381.4 $\pm$ 3.4 &  Bailer-Jones \\
\multicolumn{4}{l}{\hspace{-0.2cm} Optical and near-infrared magnitudes:}  \\
~~~$g'$ & Pan-STARRS1 $g'$ mag & $17.092 \pm 0.009$ & $15.807 \pm 0.007$ & 16.716 $\pm$ 0.018 & PS1\\
~~~$r'$ & Pan-STARRS1 $r'$ mag & $15.933 \pm 0.008$ & $14.599 \pm 0.005$ & 15.498 $\pm$ 0.011 & PS1\\
~~~$i'$ & Pan-STARRS1 $i'$ mag & $15.097 \pm 0.008$ & $13.672 \pm 0.001$ & 14.882 $\pm$ 0.018 & PS1\\
~~~$z'$ & Pan-STARRS1 $z'$ mag & $14.732 \pm 0.014$ & $13.275 \pm 0.032$ & 14.619 $\pm$ 0.011 & PS1\\
~~~$y'$ & Pan-STARRS1 $y'$ mag & $14.523 \pm 0.008$ & $13.052 \pm 0.079$ & 14.444 $\pm$ 0.009 & PS1\\
~~~$J$ & $J$ mag & ... & 11.901 $\pm$ 0.027 & ... & 2MASS\\
~~~$H$ & $H$ mag & ... & 11.270 $\pm$ 0.033 & ... & 2MASS\\
~~~$K_s$ & $K_s$ mag & ... & 11.016 $\pm$ 0.023 & ... & 2MASS\\
~~~$W1$ &  WISE1 mag & ... & 10.89 $\pm$ 0.02 & ... & WISE\\
~~~$W2$ &  WISE2 mag & ... & 10.84 $\pm$ 0.02 & ... & WISE\\
~~~$W3$ &  WISE3 mag & ... & 10.9 $\pm$ 0.07 & ... & WISE\\
~~~$G$ &  Gaia G mag & 15.7142 $\pm$ 0.0022 & 14.2970 $\pm$ 0.0022 & 15.4340 $\pm$ 0.0022 & Gaia DR3 \\
~~~$BP$ & BP mag & 16.7499 $\pm$ 0.0056 & 15.4325 $\pm$ 0.0040 & 16.3465 $\pm$ 0.0046 & Gaia DR3 \\
~~~$RP$ & RP mag & 14.7061 $\pm$ 0.0040 & 13.2556 $\pm$ 0.0031 & 14.5185 $\pm$ 0.0031 & Gaia DR3 \\
\multicolumn{4}{l}{\hspace{-0.2cm} Spectroscopic Parameters$^a$:}\\
~~~$T_{\mathrm{eff}}$ &  Effective temperature (\unit{K}) & 3711 $\pm$ 88 & 3635 $\pm$ 88 & 4230 $\pm$ 88 & This work\\
~~~$\mathrm{[Fe/H]}$ &  Metallicity (dex) & $-0.11\pm0.12$ &  $-0.09\pm0.12$ & $0.41\pm0.12$ & This work\\
~~~$\log(g)$ & Surface gravity (cgs units) & 4.74 $\pm$ 0.05 & 4.77 $\pm$ 0.05 & 4.63 $\pm$ 0.05 & This work\\
\multicolumn{4}{l}{\hspace{-0.2cm} Model-Dependent Stellar SED and Isochrone fit Parameters$^b$:}\\
~~~$M_*$ &  Mass ($M_{\odot}$) & 0.562 $\pm$ 0.024 & $0.514^{+0.025}_{-0.022}$ & 0.739 $\pm$ 0.03 & This work\\
~~~$R_*$ &  Radius ($R_{\odot}$) & $0.541\pm0.017$ & $0.489^{+0.015}_{-0.014}$ & 0.708 $\pm$ 0.022 & This work\\
~~~$L_*$ &  Luminosity ($L_{\odot}$) & $0.0529\pm0.0024$ & $0.0395 \pm 0.0015$ & 0.15$^{+0.009}_{-0.013}$ & This work\\
~~~$\rho_*$ &  Density ($\unit{g/cm^{3}}$) & $4.99^{+0.42}_{-0.37}$ & $6.19^{+0.45}_{-0.43}$ & 2.94 $^{+0.25}_{-0.22}$ & This work\\
~~~Age & Age (Gyrs) & $7.4^{+4.4}_{-4.8}$ & $7.8^{+4.2}_{-5.2}$ & 6.7 $\pm$ 4.4 & This work\\
~~~$A_v$ & Visual Absorption (mag) & $0.066^{+0.078}_{-0.048}$ & $0.131^{+0.076}_{-0.068}$ & $0.51^{+0.11}_{-0.19}$ & This work\\
\multicolumn{4}{l}{\hspace{-0.2cm} Other Stellar Parameters:}           \\
~~~$v \sin i_*$ &  Proj. Rotational Velocity (\kms{}) & $<$ 2  & $<$ 2  & $<$ 2 & This work\\
~~~$\Delta$RV &  Bulk radial Velocity (\kms{})& -12.6 $\pm$ 0.2 & 9.5 $\pm$ 0.2 & -9.2 $\pm$ 0.2 & This work\\
~~~$U, V, W$ & Gal. Vel (barycentric, \kms) & \scriptsize{-8.2$\pm$1.2,-4.5$\pm$0.7,-35.8$\pm$2.0} & \scriptsize{0.6$\pm$0.1,10.0$\pm$0.2,1.2$\pm$0.1} & \scriptsize{-31.9$\pm$0.3,-14.8$\pm$0.2,-7.0$\pm$0.1} &  This work\\
~~~$U, V, W$ & Gal. Vel (LSR, \kms)  & \scriptsize{2.9$\pm$1.4,7.8$\pm$1.0,-28.5$\pm$2.1} & \scriptsize{11.7$\pm$0.8,22.2$\pm$0.7,8.4$\pm$0.6}  & \scriptsize{-20.8$\pm$0.9,-2.6$\pm$0.7, 0.2$\pm$0.6}  & This work\\
\enddata
\tablenotetext{}{References are: Stassun \citep{stassun_tess_2018}, 2MASS \citep{cutri_2mass_2003}, Gaia DR3 \citep{gaia_collaboration_gaia_2023}, PS1 \citep{chambers_pan-starrs1_2016}, Bailer-Jones \citep{bailer-jones_estimating_2021}, Green \citep{green_3d_2019}, WISE \citep{wright_wide-field_2010}}
\tablenotetext{a}{Derived using the HPF spectral matching algorithm from \cite{stefansson_sub-neptune-sized_2020}}
\tablenotetext{b}{{\tt EXOFASTv2} derived values using MIST isochrones with the \gaia{} parallax and spectroscopic parameters in $a$) as priors.}
\end{deluxetable*}

\begin{deluxetable*}{lccccc}
\tabletypesize{\fontsize{8}{11}\selectfont}
\renewcommand\thetable{3}
\tablecaption{Summary of stellar parameters for the TOIs (continued).}
\tablehead{
\colhead{~~~Parameter}&  
\colhead{Description}&
\colhead{TOI-5414}&
\colhead{TOI-5634A}&
\colhead{TOI-5616}&
\colhead{Reference}
}
\startdata
\multicolumn{4}{l}{\hspace{-0.2cm} Main identifiers:}  \\
~~~TOI & \tess{} Object of Interest & 5414 & 5634 & 5616 & \tess{} mission \\
~~~TIC & \tess{} Input Catalogue  &  371164829 & 119585136 & 154220877 & Stassun \\
~~~2MASS & \(\cdots\)  & J07172825+3128550 & J11423454+205755 &  J12193646+4840593 & 2MASS  \\
~~~Gaia DR3 & \(\cdots\) & 886990087556438784 & 3979511431397114752 & 1545570133528453888 & Gaia DR3\\
\multicolumn{4}{l}{\hspace{-0.2cm} Equatorial Coordinates, Proper Motion and Spectral Type:} \\
~~~$\alpha_{\mathrm{J2016}}$ &  Right Ascension (RA) &  109.3678 $\pm$ 0.0312 & 175.6436 $\pm$ 0.0436 & 184.9016 $\pm$ 0.0154 & Gaia DR3\\
~~~$\delta_{\mathrm{J2016}}$ &  Declination (Dec) & 31.4818 $\pm$ 0.0278 & 20.9653 $\pm$ 0.0477 & 48.6833 $\pm$ 0.0186 & Gaia DR3\\
~~~$\mu_{\alpha}$ &  Proper motion (RA, \unit{mas/yr}) & 15.350 $\pm$ 0.039 & -60.799 $\pm$ 0.056 & -42.985 $\pm$ 0.019 & Gaia DR3\\
~~~$\mu_{\delta}$ &  Proper motion (Dec, \unit{mas/yr}) &  -22.381 $\pm$ 0.033 & -25.488 $\pm$ 0.052 & 28.531 $\pm$ 0.025 & Gaia DR3 \\
~~~$d$ &  Distance in pc  & 360.7 $\pm$ 4.3 & $322.5^{+6.2}_{-5.5}$ & 383 $\pm$ 4.2 &  Bailer-Jones \\
\multicolumn{4}{l}{\hspace{-0.2cm} Optical and near-infrared magnitudes:}  \\
~~~$g'$ & Pan-STARRS1 $g'$ mag & 16.579 $\pm$ 0.007 & 17.311 $\pm$ 0.025 & 16.869 $\pm$ 0.012  & PS1\\
~~~$r'$ & Pan-STARRS1 $r'$ mag & 15.401 $\pm$ 0.011 & 16.179 $\pm$ 0.014 & 15.713 $\pm$ 0.007  & PS1\\
~~~$i'$ & Pan-STARRS1 $i'$ mag & 14.747 $\pm$ 0.006 & 15.388 $\pm$ 0.013 & 15.059 $\pm$ 0.014  & PS1\\
~~~$z'$ & Pan-STARRS1 $z'$ mag & 14.435 $\pm$ 0.007 & 15.039 $\pm$ 0.007 & 14.775 $\pm$ 0.013  & PS1\\
~~~$y'$ & Pan-STARRS1 $y'$ mag & 14.238 $\pm$ 0.017 & 14.857 $\pm$ 0.008 & 14.625 $\pm$ 0.014  & PS1\\
~~~$J$ & $J$ mag & 13.136 $\pm$ 0.024 & 13.696 $\pm$ 0.026 & 13.566 $\pm$ 0.024 &  2MASS\\
~~~$H$ & $H$ mag &12.486 $\pm$ 0.032 &  13.071 $\pm$ 0.033 & 12.862 $\pm$ 0.021 & 2MASS\\
~~~$K_s$ & $K_s$ mag & 12.287 $\pm$ 0.025 & 12.913 $\pm$ 0.033 & 12.735 $\pm$ 0.025 &  2MASS\\
~~~$W1$ &  WISE1 mag & 12.196 $\pm$ 0.024 & 12.653 $\pm$ 0.025 & 12.587 $\pm$ 0.023 &  WISE\\
~~~$W2$ &  WISE2 mag & 12.191 $\pm$ 0.023 & 12.615 $\pm$ 0.025 & 12.862 $\pm$ 0.021 & WISE\\
~~~$W3$ &  WISE3 mag & 12.111 $\pm$ 0.38 & 12.603 $\pm$ 0.473 & 12.735 $\pm$ 0.025  & WISE\\
~~~$G$ &  Gaia G mag & 15.2958 $\pm$ 0.0025 & 15.9859 $\pm$ 0.0024 & 15.6133 $\pm$ 0.0021 & Gaia DR3 \\
~~~$BP$ & BP mag & 16.2194 $\pm$ 0.0052 & 16.9875 $\pm$ 0.0114 & 16.5177 $\pm$ 0.0039 & Gaia DR3 \\
~~~$RP$ & RP mag & 14.3592 $\pm$ 0.0036 & 14.9995 $\pm$ 0.0049 & 14.6887 $\pm$ 0.0030 & Gaia DR3 \\
\multicolumn{4}{l}{\hspace{-0.2cm} Spectroscopic Parameters$^a$:}\\
~~~$T_{\mathrm{eff}}$ &  Effective temperature (\unit{K}) & $4101\pm88$ & $3896\pm88$ & 3996 $\pm$ 88 & This work\\
~~~$\mathrm{[Fe/H]}$ &  Metallicity (dex) & $0.02\pm0.12$ & $-0.27\pm0.12$ &  $0.14\pm0.12$ &  This work\\
~~~$\log(g)$ & Surface gravity (cgs units) & $4.69\pm0.05$ & $4.73\pm0.05$ & $4.67\pm0.05$ & This work\\
\multicolumn{4}{l}{\hspace{-0.2cm} Model-Dependent Stellar SED and Isochrone fit Parameters$^b$:}\\
~~~$M_*$ &  Mass ($M_{\odot}$) & $0.706^{+0.029}_{-0.025}$ & $0.556\pm0.022$ & 0.666$\pm$0.025 &  This work\\
~~~$R_*$ &  Radius ($R_{\odot}$) & $0.691\pm0.018$ & $0.540^{+0.017}_{-0.016}$ & 0.646 $\pm$ 0.017 & This work\\
~~~$L_*$ &  Luminosity ($L_{\odot}$) & $0.1295^{+0.0061}_{-0.0054}$ & $0.0569^{+0.0028}_{-0.0024}$ & 0.1017 $\pm$ 0.0038 &  This work\\
~~~$\rho_*$ &  Density ($\unit{g/cm^{3}}$) & $3.02^{+0.23}_{-0.21}$ & $4.99^{+0.39}_{-0.36}$ & $3.49^{+0.26}_{-0.23}$ &  This work\\
~~~Age & Age (Gyrs) & $7.2^{+4.4}_{-4.6}$ & $7.4^{+4.5}_{-4.7}$ & $7.2^{+4.4}_{-4.5}$  & This work\\
~~~$A_v$ & Visual Absorption (mag) & $0.083^{+0.088}_{-0.058}$ & $0.039^{+0.032}_{-0.027}$ & 0.05 $\pm$ 0.045  & This work\\
\multicolumn{4}{l}{\hspace{-0.2cm} Other Stellar Parameters:}           \\
~~~$v \sin i_*$ &  Proj. Rotational Velocity (\kms{}) & $<$ 2 & $<$ 2 & $<$ 3 & This work\\
~~~$\Delta$RV &  Bulk radial Velocity (\kms{}) & -46.3 $\pm$ 0.2 & -22.4 $\pm$ 0.2 & -7.1 $\pm$ 0.2 & This work\\
~~~$U, V, W$ & Gal. Vel (barycentric, \kms{}) & \scriptsize{52.1$\pm$0.2,-39.5$\pm$0.8,-4.7$\pm$0.2} & \scriptsize{-57.3$\pm$1.9,-64.6$\pm$2.2,-49.6$\pm$0.9} & \scriptsize{-90.1$\pm$1.4,3.6$\pm$0.1,-35.8$\pm$0.5} &  This work\\
~~~$U, V, W$ & Gal. Vel (LSR, \kms{})  &  \scriptsize{63.2$\pm$0.9,-27.2$\pm$1.0,2.5$\pm$0.6} & \scriptsize{-46.2$\pm$2.1,-52.4$\pm$2.3, -42.3$\pm$1.1} & \scriptsize{-79.0$\pm$1.6,15.8$\pm$0.7,-28.6$\pm$0.8}  & This work\\
\enddata
\tablenotetext{}{References are: Stassun \citep{stassun_tess_2018}, 2MASS \citep{cutri_2mass_2003}, Gaia DR3 \citep{gaia_collaboration_gaia_2023}, PS1 \citep{chambers_pan-starrs1_2016}, Bailer-Jones \citep{bailer-jones_estimating_2021}, Green \citep{green_3d_2019}, WISE \citep{wright_wide-field_2010}}
\tablenotetext{a}{Derived using the HPF spectral matching algorithm from \cite{stefansson_sub-neptune-sized_2020}}
\tablenotetext{b}{{\tt EXOFASTv2} derived values using MIST isochrones with the \gaia{} parallax and spectroscopic parameters in $a$) as priors.}
\end{deluxetable*}

\section{Stellar Parameters}\label{sec:stellar}
\subsection{Spectroscopic Parameters}
For all six TOIs, we first run \texttt{HPF-SpecMatch}\footnote{\url{https://gummiks.github.io/hpfspecmatch/}} \citep{stefansson_sub-neptune-sized_2020}, to estimate \teff, ${\rm log}(g)$, and Fe/H. This is an empirical spectral matching technique which utilizes a library of HPF spectra for GKM dwarfs with well-defined labels, spanning \teff{} from 2700 K to 6000 K, ${\rm log}(g)$ from 4.3 to 5.3, and Fe/H from -0.5 to +0.5 dex, and is adapted from \texttt{SpecMatch-Emp} \citep{yee_precision_2017}. The target spectrum is compared with each star from the library to calculate a $\chi^2$, which is then used to rank the top five library stars. The spectral labels for these five stars are then combined through a linear combination weighted by their relative $\chi^2$ values. The errors on each label are estimated using a leave-one-out cross-validation process. We use HPF order index 5 for this analysis (8534 -- 8645 \AA), because it has minimal telluric contamination. Based on the $\chi^2$ for each parameter, we caution that the recovered Fe/H estimates are not robust and, while we provide their nominal estimates for completeness, we recommend their usage only in a categorical sense (super-Solar, Solar, or sub-Solar). The spectral resolution of HPF enables us to place an upper limit of $<$ 2 \kms{} for the $v \sin i_*$ of all these stars.

\subsection{Model Dependent Parameters}
We then use these spectroscopic parameters as priors to determine model-dependent stellar parameters by modelling the spectral energy distribution (SED) for the six TOIs using the \texttt{EXOFASTv2} routine \citep{eastman_exofastv2_2019}. The SED fit uses the precomputed bolometric corrections\footnote{\url{http://waps.cfa.harvard.edu/MIST/model_grids.html\#bolometric}} in \(\log g_\star\), \teff{}, [Fe/H], and \(A_V\) from the MIST model grids \citep{dotter_mesa_2016, choi_mesa_2016}. To do this, we place Gaussian priors on the photometry specified in \autoref{tab:stellarparam}, the spectroscopic parameters from \texttt{HPF-SpecMatch}, and the distance calculated from \citep{bailer-jones_estimating_2021}. The spectroscopic and derived stellar parameters for all six TOIs are given in \autoref{tab:stellarparam}. We note that due to the presence of nearby companions for K2-419A, and TOI-5218, we only use Pan-STARRS1 photometry \citep{chambers_pan-starrs1_2016} to fit the SED for those two stars, whereas for the other four we also include WISE \citep{wright_wide-field_2010} and 2MASS \citep{cutri_2mass_2003}. Given the faintness of all six TOIs, while we list the \gaia{} prism (\textit{BP, RP}) magnitudes in \autoref{tab:stellarparam}, we do not include them in the SED fits.

\subsection{UVW velocities}
We also utilize the bulk radial velocity (derived from HPF spectra), and proper motion from Gaia DR3 \citep{gaia_collaboration_gaia_2023}, to calculate the UVW velocities in the local standard of rest \citep[LSR; ][]{schonrich_local_2010} and barycentric frames using \texttt{GALPY} \citep{bovy_galpy_2015}\footnote{With \textit{U} towards the Galactic center, \textit{V} towards the direction of Galactic spin, and \textit{W} towards the North Galactic Pole \citep{johnson_calculating_1987}.}. We then use the BANYAN tool \citep{gagne_banyan_2018} to check the membership probabilities for each of our stars in the thin disk, thick disk, halo or as part of moving groups. We find that all six TOIs are likely part of the thin disk as field stars.

\subsection{Companion Stars}\label{sec:stellarcompanions}
Three of our stars (K2-419A, TOI-6034 and TOI-5634A), are part of bound pairs according to the \gaia{} catalogue from \cite{el-badry_million_2021}. K2-419A has a companion (TIC-800461642, Gaia DR3 605593554127091200) $\sim$ 2.2\arcsec~(projected separation $\sim 520$ AU) away that is resolved in \gaia{} DR3 astrometry, and at a distance of $\sim$ 224 $\pm$ 21 pc. With a $G$ mag of  19.1, it has a $\Delta G' = 3.4$ mag, and is likely a mid/late M-dwarf. Given its faintness, this companion (referred to as K2-419B from hereon) does not have a $BP - RP$ estimate. TOI-6034 also has a companion (TIC-388076435, Gaia DR3 2270407952771897472) detected in \gaia{} DR3 with similar proper motion and distance to TOI-6034, and at a separation of $\sim$ 40\arcsec~\citep[projected separation of $\sim$ 4700 A; ][]{gaia_collaboration_gaia_2023}. This is a late-F star,  with a $G$ mag of 9.58 and $BP - RP$ of 0.75, with a $\Delta G' = - 4.7$ mag, and causes significant contamination in the TESS light curves.  TOI-5634A has an early/mid comoving M-dwarf (TIC 903545876, Gaia DR3 3979511431397963008) at a separation of about 4\arcsec~(projected separation of $\sim$ 1230 AU) and $\Delta G' = 2.3$ mag. None of these three companion stars are close enough to the planet host stars to cause appreciable dilution in the HPF (1.7\arcsec~fiber diameter) or MAROON-X (0.77\arcsec~fiber diameter) fiber-fed spectra.

TOI-6034 is the first GEMS host to exist as part of a wide-separation binary with an earlier (FGK) main sequence companion, while the other two join the growing list of GEMS existing in binary systems including HATS-74 A b \citep{jordan_hats-74ab_2022}, TOI-3714 b \citep{canas_toi-3714_2022}, TOI-3984 A b \citep{canas_toi-3984_2023}, TOI-5293 A b \citep{canas_toi-3984_2023}, and TOI-6383 A b (Marta Bernab\`o et al., submitted). We will contextualize the binary separation and binary fraction of GEMS and its potential impact of giant planet formation in a future work.

\begin{deluxetable*}{llcc}
\tablecaption{Derived Parameters for the K2-419A and TOI-6034 System.   \label{tab:planetprop}}
\tablehead{\colhead{~~~Parameter} &
\colhead{Units} &
\colhead{K2-419A$^a$} & 
\colhead{TOI-6034$^a$} 
}
\startdata
\sidehead{Orbital Parameters:}
~~~Orbital Period\dotfill & $P$ (days) \dotfill & 20.35847252$^{+0.00000614}_{-0.00000575}$ & 2.576184 $\pm$ 0.000002\\
~~~Eccentricity\dotfill & $e$ \dotfill & 0.042$^{+0.045}_{-0.029}$; (e$_{< 97.5\%}$ $<$ 0.147) & 0.040$^{+0.042}_{-0.028}$; (e$_{< 97.5\%}$ $<$ 0.133)  \\
~~~Argument of Periastron\dotfill & $\omega$ (radians) \dotfill & 0.017$^{+1.022}_{-1.214}$ & -0.59$^{+1.75}_{-1.31}$ \\
~~~Semi-amplitude Velocity\dotfill & $K$ (\ms{})\dotfill & 67$^{+5}_{-4}$ & 184 $\pm$ 16\\
 ~~~Systemic Velocity$^b$\dotfill & $\gamma_{\mathrm{HPF}}$ (\ms{})\dotfill & -836 $\pm$ 54 & 6 $\pm$ 15\\
~~~ & $\gamma_{\mathrm{M-X Blue}}$ (\ms{})\dotfill & -16 $\pm$ 24 & ...\\
~~~ & $\gamma_{\mathrm{M-X Red}}$ (\ms{})\dotfill & -19 $\pm$ 23 & ...\\
~~~RV trend\dotfill & $dv/dt$ (\ms{} yr$^{-1}$)   & 27.28$^{+66.43}_{-67.20}$ & -15.5$^{+82.1}_{-85.6}$   \\ 
~~~ & $dv/dt^2$ (\ms{} yr$^{-2}$)   & 37.40$^{+69.42}_{-71.57}$ & -0.5 $\pm$ 95.5   \\ 
~~~RV jitter\dotfill & $\sigma_{\mathrm{HPF}}$ (\ms{})\dotfill & 49.3$^{+32.5}_{-33.0}$ & 35.5$^{+15.9}_{-11.9}$\\
~~~ & $\sigma_{\mathrm{MX~Blue}}$ (\ms{})\dotfill & 18.4$^{+7.8}_{-5.8}$ & ...\\
~~~ & $\sigma_{\mathrm{MX~Red}}$ (\ms{})\dotfill & 6.7$^{+4.7}_{-3.9}$ & ...\\
\sidehead{Transit Parameters:}
~~~Transit Midpoint \dotfill & $T_C$ (BJD\textsubscript{TDB})\dotfill & 2459553.7138 $\pm$ 0.0005 & 2459883.9055 $\pm$ 0.0003 \\
~~~Scaled Radius\dotfill & $R_{p}/R_{*}$ \dotfill & 
0.1805$^{+0.0031}_{-0.0028}$ & 0.2233$^{+0.0051}_{-0.0052}$ \\
~~~Scaled Semi-major Axis\dotfill & $a/R_{*}$ \dotfill & 48.46$^{+1.53}_{-1.40}$ & 12.95$^{+0.41}_{-0.37}$\\
~~~Impact Parameter\dotfill & $b$ & 0.389$^{+0.075}_{-0.103}$ & 0.674$^{+0.035}_{-0.040}$ \\
~~~Orbital Inclination\dotfill & $i$ (degrees)\dotfill & 89.54$^{+0.12}_{-0.09}$ & 87.06$^{+0.22}_{-0.21}$\\
~~~Transit Duration\dotfill & $T_{14}$ (days)\dotfill & 0.1489$^{+0.0041}_{-0.0037}$ & 0.0648$^{+0.0019}_{-0.0020}$\\
\sidehead{Planetary Parameters:}
~~~Mass\dotfill & $M_{p}$ (M$_\oplus$)\dotfill &  196 $\pm$ 15 & 254 $\pm$ 24 \\
~~~ & $M_{p}$ ($M_J$)\dotfill &  0.617 $\pm$ 0.047 & 0.798 $\pm$ 0.075 \\
~~~Radius\dotfill & $R_{p}$  (R$_\oplus$) \dotfill& $10.55\pm0.39$ & 11.92 $\pm$ 0.48 \\
~~~ & $R_{p}$  ($R_J$) \dotfill& 0.941 $\pm$ 0.034 & 1.063 $\pm$ 0.042 \\
~~~Density\dotfill & $\rho_{p}$ (\gcmcubed{})\dotfill & 0.916$^{+0.136}_{-0.114}$ & $0.82^{+0.14}_{-0.12}$\\
~~~Semi-major Axis\dotfill & $a$ (AU) \dotfill & 0.1208$^{+0.0016}_{-0.0017}$ & 0.02949$^{+0.00043}_{-0.00044}$\\
~~~Average Incident Flux$^c$\dotfill & $\langle F \rangle$ (\unit{10^5\ W/m^2})\dotfill &  0.045 $\pm$ 0.005 & 0.587 $\pm$ 0.068 \\
~~~Planetary Insolation & $S$ (S$_\oplus$)\dotfill &  $3.3\pm0.4$ & 43 $\pm$ 5 \\
~~~Equilibrium Temperature$^d$ \dotfill & $T_{\mathrm{eq}}$ (K)\dotfill & 377 $\pm$ 11 &  714 $\pm$ 21\\
\enddata
\tablenotetext{a}{The reported values refer to the 16-50-84\% percentile of the posteriors.}
\tablenotetext{b}{In addition to the ``Absolute RV" from \autoref{tab:stellarparam}.}

\tablenotetext{c}{We use a Solar flux constant = 1360.8 W/m$^2$, to convert insolation to incident flux.}
\tablenotetext{d}{We assume the planet to be a black body with zero albedo and perfect energy redistribution to estimate the equilibrium temperature. }
\normalsize
\end{deluxetable*}

\begin{deluxetable*}{llcccc}
\tabletypesize{\fontsize{7}{10}\selectfont}
\tablecaption{Derived Parameters for the TOI-5414, TOI-5616, TOI-5634A~And TOI-5218 Systems.   \label{tab:planetprop2}}
\tablehead{\colhead{~~~Parameter} &
\colhead{Units} &
\colhead{TOI-5414} & 
\colhead{TOI-5616} &
\colhead{TOI-5634A} & 
\colhead{TOI-5218}
}
\startdata
\sidehead{Orbital Parameters:}
~~~Orbital Period\dotfill & $P$ (days) \dotfill & 10.778918$^{+0.000089}_{-0.000092}$ & 2.002849$ \pm 0.000004$ & 2.2035373 $\pm$ 0.0000084 & 4.291452 $\pm$ 0.000004 \\
~~~Eccentricity\dotfill & $e$ \dotfill & \multicolumn{4}{c}{0 (fixed)} \\
~~~Argument of Periastron\dotfill & $\omega$ (degrees) \dotfill &  \multicolumn{4}{c}{90 (fixed)} \\
~~~Semi-amplitude Velocity\dotfill & $K$ (\ms{})\dotfill & 51$^{+35}_{-32}$ &  161$^{+73}_{-77}$ & 137$^{+96}_{-82}$ & 121$^{+59}_{-54}$ \\
~~~ & ${<97.5\%}$ ; ${<99.85\%}$ (\ms) & $<$ 122; $<$ 164 & $<$ 305; $<$ 384 & $<$ 330; $<$ 439 & $<$ 242; $<$ 309 \\
~~~Systemic Velocity$^a$\dotfill & $\gamma_{\mathrm{HPF}}$ (\ms{})\dotfill & 114 $\pm$ 35 & -188 $\pm$ 50 & -380 $\pm$ 77 & -515 $\pm 42$\\
~~~RV trend\dotfill & $dv/dt$ (\ms{} yr$^{-1}$)   &  85 $\pm$ 42  & 92 $\pm$ 58 & -29 $\pm$ 97 & -0.5 $\pm$ 99.4\\ 
~~~ & $dv/dt^2$ (\ms{} yr$^{-2}$)   &  -30 $^{+83}_{-85}$  & 62 $\pm$ 93 & -2 $\pm$ 95 & 0.5 $\pm 100$ \\ 
~~~RV jitter\dotfill & $\sigma_{\mathrm{HPF}}$ (\ms{})\dotfill & 96$^{+3}_{-6}$ & 89$^{+8}_{-14}$ & 49 $\pm$ 34 & 46 $\pm$ 34 \\
\sidehead{Transit Parameters:}
~~~Transit Midpoint \dotfill & $T_C$ (BJD\textsubscript{TDB})\dotfill & 2459529.1387$^{+0.0028}_{-0.0025}$ & 2458901.48892 $\pm$ 0.00129 & 2459663.4184 $\pm$ 0.0011 & 2458842.65207$^{+0.00076}_{-0.00079}$\\
~~~Scaled Radius\dotfill & $R_{p}/R_{*}$ \dotfill & 0.1333$^{+0.0136}_{-0.0125}$ & 0.1532$^{+0.0047}_{-0.0044}$ & 0.177 $\pm$ 0.006  & 0.130 $\pm$ 0.002 \\
~~~Scaled Semi-major Axis\dotfill & $a/R_{*}$ \dotfill & 26.29$^{+0.73}_{-0.71}$ & 10.21$^{+0.32}_{-0.34}$ & 10.85$^{+0.41}_{-0.44}$ & 14.56 $\pm$ 0.43 \\
~~~Impact Parameter\dotfill & $b$ & 0.21$^{+0.17}_{-0.15}$ & 0.34$^{+0.12}_{-0.18}$ & 0.36$^{+0.12}_{-0.18}$ & 0.40$^{+0.06}_{-0.08}$\\
~~~Orbital Inclination\dotfill & $i$ (degrees)\dotfill & 89.54$^{+0.32}_{-0.37}$ & 88.13$^{+1.05}_{-0.73}$ & 88.19$^{+0.94}_{-0.69}$ & 88.41$^{+0.35}_{-0.30}$ \\
~~~Transit Duration\dotfill & $T_{14}$ (days)\dotfill & 0.1441$^{+0.0049}_{-0.0051}$ & 0.0690 $\pm$ 0.0019 & 0.0697$^{+0.0015}_{-0.0014}$ & 0.0991$^{+0.0013}_{-0.0012}$\\
\sidehead{Planetary Parameters:}
~~~Mass\dotfill & $M_{p}$ (M$_\oplus$)\dotfill &  140$^{+96}_{-87}$ & 255 $^{+118}_{-122}$ & 183 $^{+129}_{-110}$ & 254$^{+125}_{-114}$   \\
~~~ & ${<97.5\%}$ ; ${<99.85\%}$ (\earthmass) & $<$ 336; $<$ 447 & $<$ 485; $<$ 624 & $<$440 ; $<$ 588 & $<$ 511; $<$ 657 \\
~~~Radius\dotfill & $R_{p}$  (R$_\oplus$) \dotfill & 10.10$^{+1.07}_{-0.96}$ & 10.78 $^{+0.54}_{-0.47}$ & 10.42$^{+0.62}_{-0.55}$ & 9.86 $\pm$ 0.39 \\
~~~ & $R_{p}$  ($R_J$) \dotfill & 0.901$^{+0.095}_{-0.085}$ & 0.961$^{+0.048}_{-0.042}$ & 0.923$^{+0.056}_{-0.050}$ & 0.879 $\pm$ 0.034 \\
~~~Semi-major Axis\dotfill & $a$ (AU) \dotfill & 0.0850$^{+0.0010}_{-0.0011}$ & 0.0272 $\pm$ 0.0003 & 0.0272 $\pm$ 0.0003 & 0.0469 $\pm$ 0.0006 \\
~~~Average Incident Flux$^b$\dotfill & $\langle F \rangle$ (\unit{10^5\ W/m^2})\dotfill & 0.231 $\pm$ 0.023 & 1.87 $\pm$ 0.08 & 1.05 $\pm$ 0.05 & 0.85 $\pm$ 0.09 \\
~~~Planetary Insolation & $S$ (S$_\oplus$)\dotfill & 17.00 $\pm$ 1.75 & 137 $\pm$ 6 & 77 $\pm$ 5 & 62 $\pm$ 6 \\
~~~Equilibrium Temperature$ce$ \dotfill & $T_{\mathrm{eq}}$ (K)\dotfill & 566 $\pm$ 15 & 939 $\pm$ 27 & 837 $\pm$ 34 & 783 $\pm$ 21\\
\enddata
\tablenotetext{a}{In addition to the ``Absolute RV" from \autoref{tab:stellarparam}.}

\tablenotetext{b}{We use a Solar flux constant = 1360.8 W/m$^2$, to convert insolation to incident flux.}
\tablenotetext{c}{We assume the planet to be a black body with zero albedo and perfect energy redistribution to estimate the equilibrium temperature. }
\normalsize
\end{deluxetable*}

\section{Analysis}\label{sec:analysis} 

\subsection{Validation}\label{sec:validation}
For all six systems, we perform a host star density check utilizing the transit duration estimate and find them consistent with the derived host star density at 1-$\sigma$ based on the stellar parameters \citep{seager_unique_2003}. Additionally, while we obtained high-contrast imaging for all our targets, given the lack of mass measurements for four of them, we apply statistical validation techniques to the four planet candidates (TOI-5218b, TOI-5414b, TOI-5616b, TOI-5634Ab). We use the \texttt{TRICERATOPS} \citep{2020ascl.soft02004G, giacalone_vetting_2021} library to statistically validate these a subset of these candidates. Based on the PSF for ground-based transits for these candidates, we limit the search radius to $\sim$ 3\arcsec, run 50 trials for each candidate with 1 million instances for each scenario, and include the high contrast imaging curve as well as constraints from \texttt{MOLUSC} \citep{wood_characterizing_2021} utilizing the Gaia RUWE, HPF RVs and high contrast imaging curves. Since we have ground-based transits for each of these candidates, we set the Nearby False Positive Probability (NFPP) to 0, i.e., the probability that the transit signal is from a nearby resolved star. Additionally, based on the spectra and RVs for each of these objects we rule out the SB2 (spectroscopic binary) scenario, and set the probability of EB and EBx2P (EBs at twice the period) to 0.  We then use the TESS light curve after detrending with Gaussian Processes (GP, described in the next section), and obtain a False Positive Probability (FPP) for TOI-5616b of 0.89$^{+0.20}_{-0.10}$\%, for TOI-5634Ab we get 0.16$^{+0.24}_{-0.05}$\%. For TOI-5218b we use the partial ARCTIC transit from 2023 May 14 and get an FPP consistent with 0 at machine precision. For TOI-5414b, we use the 0.4 m LCO photometry, which gives us an FPP of 1.9$^{+0.4}_{-0.3}$\%. Based on the FPP recommendation from \cite{giacalone_vetting_2021}, we validate TOI-5218b, TOI-5616b and TOI-5634Ab as planets (FPP $<$ 1.5\%), and TOI-5414b is likely be a planet (FPP $<$ 50\%) but not fully validated.

\begin{figure*}[!t]
\centering
\fig{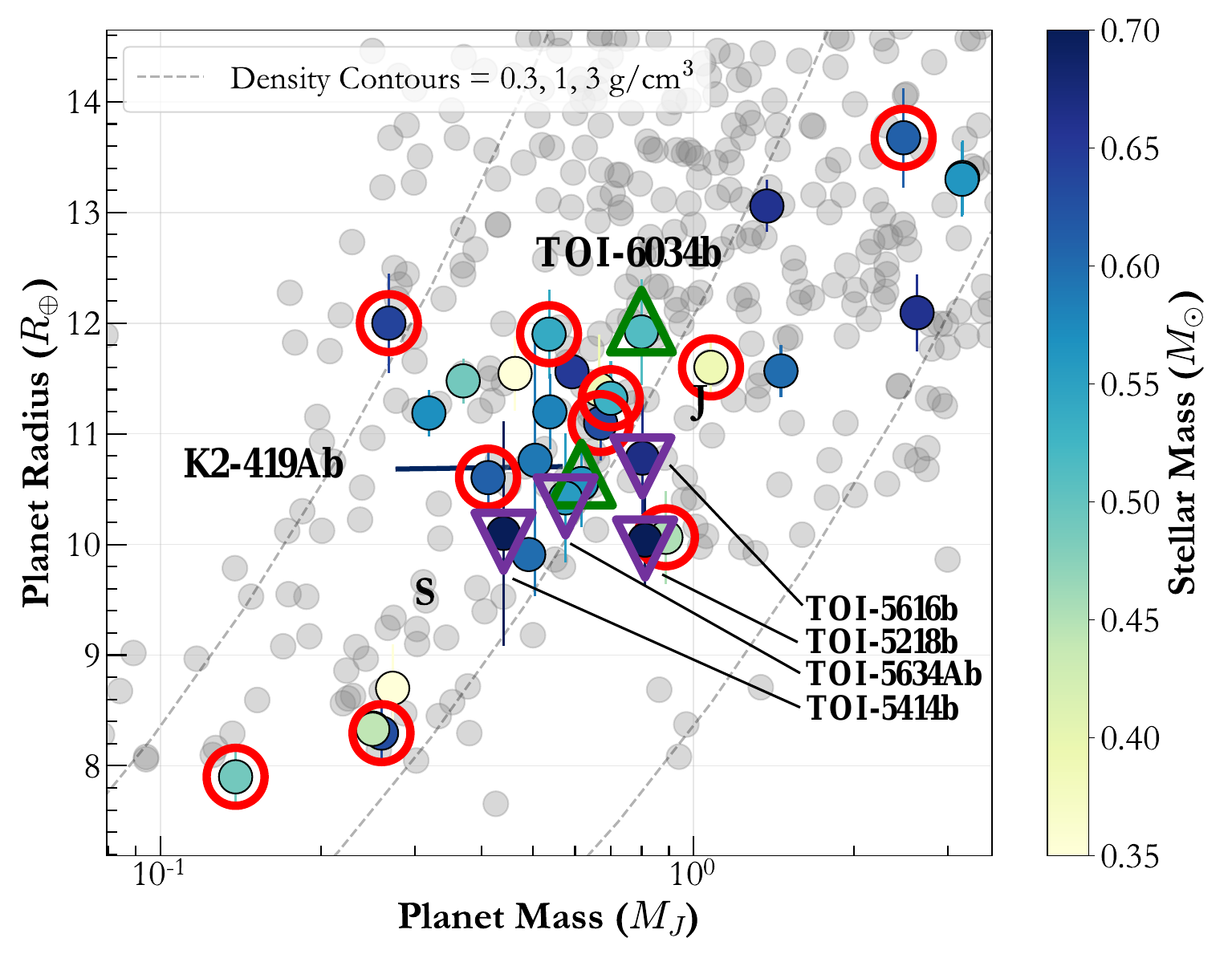}{\columnwidth}
{\small a) Planet radius as a function of mass}    \label{fig:RadiusMass}
\fig{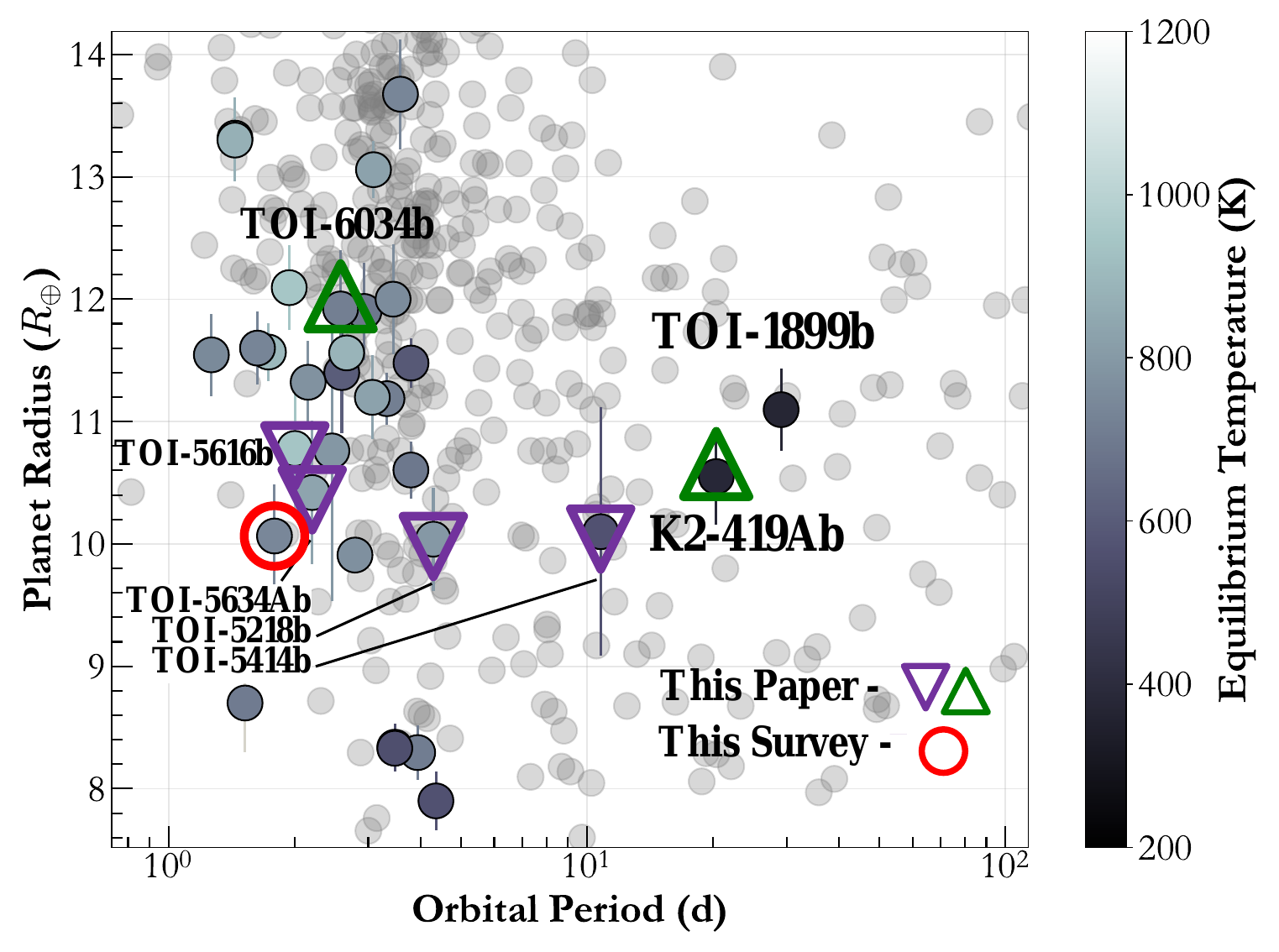}{\columnwidth}
{\small b) Planet radius as a function of orbital period} \label{fig:RadiusPeriod}

\caption{\small We show a sample of planets with host star \teff{} $<$ 7200 K, and planetary radii between 8 -- 15 \earthradius. In \textbf{a)} and \textbf{b)} we highlight systems with \teff{} $<$ 4300 K to focus on the cooler host stars. In all these plots, the systems discovered as part of the \textit{Searching for GEMS} are circled in red, while the TOIs confirmed here with triangles in green (upright for TOIs with mass measurements) and purple (inverted for TOIs with mass upper limits)}.\label{fig:ParameterSpace}
\end{figure*}

\subsection{Joint Fitting of Photometry and RVs}\label{sec:joint}
We use the \texttt{exoplanet} \citep{foreman-mackey_exoplanet_2021, foreman-mackey_exoplanet-devexoplanet_2021} package to jointly fit the photometry with RVs. This uses \texttt{pymc3} \citep{salvatier_probabilistic_2016} to perform Hamiltonian Monte-Carlo (HMC) posterior estimation, \texttt{starry} \citep{luger_starry_2019, agol_analytic_2020} to model the transits, and is based on the analytical models from \cite{mandel_analytic_2002}. We use separate quadratic limb-darkening terms for each instrument. The TESS and K2 photometric datasets are detrended using a GP \texttt{RotationTerm} kernel, which is a combination of two simple harmonic oscillators and implemented in \texttt{celerite2} \citep{foreman-mackey_fast_2017, foreman-mackey_scalable_2018}. While fitting the GP simultaneously with the transit and RV models would be ideal to prevent over-fitting and provide unbiased estimates, this adds substantial computational costs. Therefore, we use a simultaneous GP term for the TESS photometry for TOI-5616 and TOI-5634, which have fewer sectors. For the other TOIs we mask the transits and a 6 hour window on either side of each transit midpoint, and then fit a GP trend to this out-of-transit baseline, that is subsequently subtracted. This detrended light curve is then used for joint photometry + RV fitting. Due to TESS' large pixel and PSF size, we include a dilution term for each TESS sector that has a uniform prior between 0.001 and 1.5 \citep[explained in][]{kanodia_toi-5205b_2023}, and is determined using the other photometric datasets where the stars are spatially well-resolved.  For each RV instrument time series, we also include an RV jitter (white-noise) and RV offset term while combining across instruments. For each system, we include a quadratic RV trend as a function of time ($t\cdot dv/dt$ + $t^2 \cdot dv/dt^2$), where $t = 0$ is the mid-epoch of the RV time series.

For the two systems where we have mass measurements --- K2-419Ab (\autoref{fig:Data5176}) and TOI-6034b (\autoref{fig:Data6034}), we perform a full joint fit allowing for eccentricity to float. For the other four systems with mass upper limits --- TOI-5218b (\autoref{fig:Data5218}), TOI-5414b (\autoref{fig:Data5414}), TOI-5616b (\autoref{fig:Data5616} and TOI-5634Ab (\autoref{fig:Data5634}) --- we fix eccentricity to be circular given the few RV measurements and short orbital periods. The final derived parameters for all six systems are reported in \autoref{tab:planetprop} and \autoref{tab:planetprop2}.

\section{Discussion}\label{sec:discussion}

TESS has contributed to the discovery of many short-period giant exoplanets with targeted surveys such as our \textit{Searching for GEMS} survey \citep{kanodia_searching_2024} geared towards M-dwarf hosts, and the \textit{TESS Grand Unified Hot Jupiter Survey} \citep{2022AJ....164...70Y} focussed on nearby brighter FGK stars. This has led to an increasing number of giant planets orbiting brighter hosts, more amenable to detailed characterization, compared to the \textit{Kepler} mission.

In particular, the sample of confirmed transiting GEMS has grown from less than a handful to about 20 confirmed planets, with many more being added by our survey. The six planets discussed in this manuscript are contextualized with respect to the existing sample of planets \citep{ps} in \autoref{fig:ParameterSpace}. We discuss some salient features of the systems here.

\textbf{K2-419Ab} is one of the longest period GEMS ($\sim$ 20 days; \autoref{fig:ParameterSpace}) after TOI-1899~b \citep[$\sim$ 29 days;][]{canas_warm_2020, lin_unusual_2023}. Also similar to TOI-1899~b, despite having a large  $a/R_*$  of $\sim$ 48, its orbit is consistent with circular at 1.5 $\sigma$. This agrees with the behaviour seen by \cite{dawson_giant_2013}, where metal-rich stars tend to host warm/cold Jupiters spanning a range of eccentricities, whereas relatively metal-poorer host stars tend to harbour more low-eccentricity warm Jupiters. K2-419A does not have rotational broadening or a rotation signal in the spectroscopic activity indicators. This conforms with studies of hot/warm/cold Jupiter samples by \cite{banerjee_host_2024}, which find that warm and cold Jupiters are preferentially around old and relatively metal-poor stars. 

The equilibrium temperature for K2-419Ab at $\sim$ 380~K also makes it one of the coolest giant planets with precise mass and radius measurements. In particular, planets cooler than 500 K are conducive for detection of ammonia \citep[NH$_3$;][]{ohno_nitrogen_2023}, which can be used a tracer for the Nitrogen abundance in the planetary atmosphere. In combination with carbon measured using carbon-bearing species such as carbon mono-oxide and Methane, estimates of carbon, oxygen and nitrogen can be used to make a first attempt towards reconciling the volatile content of mature exoplanet atmospheres with theories which predict volatile depletion \citep{lissauer_planets_2007, ciesla_volatile_2015} and recent JWST MIRI protoplanetary disk observations; where the latter find a diverse set of conditions at these orbital separations in M-dwarf disks \citep{van_dishoeck_diverse_2023, franceschi_minds_2024}.  A similar temperature around a Solar analogue necessitates a separation of 0.55 AU from its host star (or a period of $\sim$ 150 days), which hampers transit observability and results in a long transit duration.

\textbf{TOI-5414b} (and \textbf{TOI-5616b}), suggest the presence of a linear (and quadratic) RV trend (\autoref{tab:planetprop2}). However, given the large uncertainties in the RV errors for these faint stars, as well as the large posterior width of the inferred RV trend, we advise caution in interpreting these potential trends.

\section{Summary}\label{sec:conclusion}
In this manuscript, we describe the observations and analysis utilized to confirm five planets, which included mass measurements for two planets (K2-419Ab and TOI-6034b), statistical validation and upper limits for three TOIs (TOI-5218b, TOI-5616b, and TOI-5634Ab), and the classification of TOI-5414b as a likely planet. Two of these are mid/late K-dwarfs (TOI-5414b and TOI-5218b), while the rest are GEMS as part of our \textit{Searching for GEMS} survey. Interestingly, of the four GEMS TOIs, three form part of binary systems, which adds to the growing sample of confirmed GEMS that exist in binary systems. K2-419Ab, which was observed with K2 and TESS photometry, along with HPF and MAROON-X radial velocities is one of the coolest transiting giant planets with precise characterization. TOI-6034 hosts a transiting giant planet, and is the only one so far to have an early (F-type) main-sequence binary companion. These two planets, along with the other four TOIs with mass upper limits contribute to future statistical analysis to ascertain the possible presence of a trend between giant planet mass and stellar mass.

\section{Acknowledgements}

SK would like to thank the referee for their detailed comments and suggestions which have helped improve this manuscript.

SK would like to thank Peter Gao for help with the computing resources that enabled running some of these memory intensive analyses.

CIC acknowledges support by NASA Headquarters through an appointment to the NASA Postdoctoral Program at the Goddard Space Flight Center, administered by ORAU through a contract with NASA.

The Pennsylvania State University campuses are located on the original homelands of the Erie, Haudenosaunee (Seneca, Cayuga, Onondaga, Oneida, Mohawk, and Tuscarora), Lenape (Delaware Nation, Delaware Tribe, Stockbridge-Munsee), Shawnee (Absentee, Eastern, and Oklahoma), Susquehannock, and Wahzhazhe (Osage) Nations.  As a land grant institution, we acknowledge and honor the traditional caretakers of these lands and strive to understand and model their responsible stewardship. We also acknowledge the longer history of these lands and our place in that history.

These results are based on observations obtained with the Habitable-zone Planet Finder Spectrograph on the HET. We acknowledge support from NSF grants AST-1006676, AST-1126413, AST-1310885, AST-1310875,  ATI 2009889, ATI-2009982, AST-2108512, AST-2108801 and the NASA Astrobiology Institute (NNA09DA76A) in the pursuit of precision radial velocities in the NIR. The HPF team also acknowledges support from the Heising-Simons Foundation via grant 2017-0494. The Low Resolution Spectrograph 2 (LRS2) was developed and funded by the University of Texas at Austin McDonald Observatory and Department of Astronomy and by Pennsylvania State University. We thank the Leibniz-Institut für Astrophysik Potsdam (AIP) and the Institut für Astrophysik Göttingen (IAG) for their contributions to the construction of the integral field units.  The Hobby-Eberly Telescope is a joint project of the University of Texas at Austin, the Pennsylvania State University, Ludwig-Maximilians-Universität München, and Georg-August Universität Gottingen. The HET is named in honor of its principal benefactors, William P. Hobby and Robert E. Eberly. The HET collaboration acknowledges the support and resources from the Texas Advanced Computing Center. We thank the Resident astronomers and Telescope Operators at the HET for the skillful execution of our observations with HPF. We would like to acknowledge that the HET is built on Indigenous land. Moreover, we would like to acknowledge and pay our respects to the Carrizo \& Comecrudo, Coahuiltecan, Caddo, Tonkawa, Comanche, Lipan Apache, Alabama-Coushatta, Kickapoo, Tigua Pueblo, and all the American Indian and Indigenous Peoples and communities who have been or have become a part of these lands and territories in Texas, here on Turtle Island.

Observations obtained with the Apache Point Observatory 3.5-meter telescope, owned and operated by the Astrophysical Research Consortium. 
We acknowledge support from NSF grants AST-1910954, AST-1907622, AST-1909506, AST-1909682 for the ultra-precise photometry effort.

This work makes use of observations (Proposal ID: KEY2020B-005) from the Sinistro imaging camera on the 1m Dome B telescope at Cerro Tololo Inter-American Observatory, operated by the Las Cumbres Observatory global telescope network (LCOGT).

DRC and CAC acknowledge partial support from NASA Grant 18-2XRP18\_2-0007. This research has made use of the Exoplanet Follow-up Observation Program (ExoFOP; DOI: 10.26134/ExoFOP5) website, which is operated by the California Institute of Technology, under contract with the National Aeronautics and Space Administration under the Exoplanet Exploration Program. The authors wish to recognize and acknowledge the very significant cultural role and reverence that the summit of Mauna Kea has always had within the Native Hawaiian community. We are most fortunate to have the opportunity to conduct observations from this mountain.

CAC and SPH acknowledge that this research was carried out at the Jet Propulsion Laboratory, California Institute of Technology, under a contract with the National Aeronautics and Space Administration (80NM0018D0004).

WIYN is a joint facility of the University of Wisconsin-Madison, Indiana University, NSF's NOIRLab, the Pennsylvania State University, Purdue University, University of California-Irvine, and the University of Missouri. 
The authors are honored to be permitted to conduct astronomical research on Iolkam Du'ag (Kitt Peak), a mountain with particular significance to the Tohono O'odham. Data presented herein were obtained at the WIYN Observatory from telescope time allocated to NN-EXPLORE (PI: Gupta; 2022A-665981, 2022B-936991, 2023A-845810, PI-Kanodia; 2023B-438370, 2024A-103024) through the scientific partnership of NASA, the NSF, and NOIRLab. 

Deepest gratitude to Zade Arnold, Joe Davis, Michelle Edwards, John Ehret, Tina Juan, Brian Pisarek, Aaron Rowe, Fred Wortman, the Eastern Area Incident Management Team, and all of the firefighters and air support crew who fought the recent Contreras fire. Against great odds, you saved Kitt Peak National Observatory.

Some of the observations in this paper made use of the NN-EXPLORE Exoplanet and Stellar Speckle Imager (NESSI). NESSI was funded by the NASA Exoplanet Exploration Program and the NASA Ames Research Center. NESSI was built at the Ames Research Center by Steve B. Howell, Nic Scott, Elliott P. Horch, and Emmett Quigley.


Some of the observations in the paper made use of the High-Resolution Imaging instrument(s) `Alopeke. `Alopeke was funded by the NASA Exoplanet Exploration Program and built at the NASA Ames Research Center by Steve B. Howell, Nic Scott, Elliott P. Horch, and Emmett Quigley. `Alopeke was mounted on the Gemini North (and/or South) telescope of the international Gemini Observatory, a program of NSF NOIRLab, which is managed by the Association of Universities for Research in Astronomy (AURA) under a cooperative agreement with the U.S. National Science Foundation on behalf of the Gemini Observatory partnership: the U.S. National Science Foundation (United States), National Research Council (Canada), Agencia Nacional de Investigaci\'{o}n y Desarrollo (Chile), Ministerio de Ciencia, Tecnolog\'{i}a e Innovaci\'{o}n (Argentina), Minist\'{e}rio da Ci\^{e}ncia, Tecnologia, Inova\c{c}\~{o}es e Comunica\c{c}\~{o}es (Brazil), and Korea Astronomy and Space Science Institute (Republic of Korea).

The University of Chicago group acknowledges funding for the MAROON-X project from the David and Lucile Packard Foundation, the Heising-Simons Foundation, the Gordon and Betty Moore Foundation, the Gemini Observatory, the NSF (award number 2108465), and NASA (grant number 80NSSC22K0117). The Gemini observations are associated with program GN-2023B-Q-104 (PI: Kanodia)

This material is based upon work supported by the National Science Foundation Graduate Research Fellowship under Grant No. DGE 1746045

This work has made use of data from the European Space Agency (ESA) mission Gaia (\url{https://www.cosmos.esa.int/gaia}), processed by the Gaia Data Processing and Analysis Consortium (DPAC, \url{https://www.cosmos.esa.int/web/gaia/dpac/consortium}). Funding for the DPAC has been provided by national institutions, in particular the institutions participating in the Gaia Multilateral Agreement.

Some of the observations in this paper were obtained with the Samuel Oschin Telescope 48-inch and the 60-inch Telescope at the Palomar Observatory as part of the ZTF project. ZTF is supported by the NSF under Grant No. AST-2034437 and a collaboration including Caltech, IPAC, the Weizmann Institute for Science, the Oskar Klein Center at Stockholm University, the University of Maryland, Deutsches Elektronen-Synchrotron and Humboldt University, the TANGO Consortium of Taiwan, the University of Wisconsin at Milwaukee, Trinity College Dublin, Lawrence Livermore National Laboratories, and IN2P3, France. Operations are conducted by COO, IPAC, and UW.

Computations for this research were performed on the Pennsylvania State University’s Institute for Computational and Data Sciences Advanced CyberInfrastructure (ICDS-ACI).  This content is solely the responsibility of the authors and does not necessarily represent the views of the Institute for Computational and Data Sciences.

The Center for Exoplanets and Habitable Worlds is supported by the Pennsylvania State University, the Eberly College of Science, and the Pennsylvania Space Grant Consortium. 

Some of the data presented in this paper were obtained from MAST at STScI. Support for MAST for non-HST data is provided by the NASA Office of Space Science via grant NNX09AF08G and by other grants and contracts.

This work includes data collected by the TESS mission, which are publicly available from MAST. Funding for the TESS mission is provided by the NASA Science Mission directorate. 
This research made use of the (i) NASA Exoplanet Archive, which is operated by Caltech, under contract with NASA under the Exoplanet Exploration Program, (ii) SIMBAD database, operated at CDS, Strasbourg, France, (iii) NASA's Astrophysics Data System Bibliographic Services, and (iv) data from 2MASS, a joint project of the University of Massachusetts and IPAC at Caltech, funded by NASA and the NSF.

This research has made use of the SIMBAD database, operated at CDS, Strasbourg, France, 
and NASA's Astrophysics Data System Bibliographic Services.

This research has made use of the Exoplanet Follow-up Observation Program website, which is operated by the California Institute of Technology, under contract with the National Aeronautics and Space Administration under the Exoplanet Exploration Program

CIC acknowledges support by an appointment to the NASA Postdoctoral Program at the Goddard Space Flight Center, administered by USRA through a contract with NASA.

\facilities{\gaia{}, HET (HPF), WIYN 3.5 m (NESSI), Shane (ShARCS), Gemini-N (MAROON-X), , Gemini-N (`Alopeke), Palomar AO (PHARO), 1.2~m FLWO (KeplerCam) , Swope 1.0 m, RBO, Keeble, LCOGT 0.4 m Teide, ZTF, \tess{}, K2, LCRO 0.3 m, Exoplanet Archive}
\software{
\texttt{ArviZ} \citep{kumar_arviz_2019}, 
AstroImageJ \citep{collins_astroimagej_2017}, 
\texttt{astrometry.net} \citep{hogg_automated_2008},
\texttt{astroquery} \citep{ginsburg_astroquery_2019}, 
\texttt{astropy} \citep{robitaille_astropy_2013, astropy_collaboration_astropy_2018},
BANYAN \citep{gagne_banyan_2018},
BANZAI \citep{mccully_real-time_2018},
\texttt{barycorrpy} \citep{kanodia_python_2018}, 
\texttt{celerite2} \citep{foreman-mackey_fast_2017, foreman-mackey_scalable_2018},
\texttt{DEATHSTAR} \citep{ross_deathstar_2023},
\texttt{eleanor} \citep{feinstein_eleanor_2019},
\texttt{EVEREST} \citep{luger_everest_2016, luger_update_2018},
\texttt{EXOFASTv2} \citep{eastman_exofastv2_2019},
\texttt{exoplanet} \citep{foreman-mackey_exoplanet-devexoplanet_2021, foreman-mackey_exoplanet_2021},
\texttt{HPF-SpecMatch} \citep{stefansson_sub-neptune-sized_2020},
\texttt{HxRGproc} \citep{ninan_habitable-zone_2018},
\texttt{GALPY} \citep{bovy_galpy_2015},
\texttt{ipython} \citep{perez_ipython_2007},
\texttt{lightkurve} \citep{lightkurve_collaboration_lightkurve_2018},
\texttt{matplotlib} \citep{hunter_matplotlib_2007},
\texttt{MOLUSC} \citep{wood_characterizing_2021},
\texttt{numpy} \citep{oliphant_numpy_2006},
\texttt{pandas} \citep{mckinney_data_2010},
\texttt{photutils} \citep{bradley_astropyphotutils_2020},
\texttt{pyastrotools} \citep{kanodia_shbhukpyastrotools_2023},
\texttt{PyMC3} \citep{salvatier_probabilistic_2016},
\texttt{scipy} \citep{oliphant_python_2007, virtanen_scipy_2020},
\texttt{SERVAL} \citep{zechmeister_spectrum_2018},
\texttt{tglc} \citep{han_tess-gaia_2023},
\texttt{Theano} \citep{the_theano_development_team_theano_2016},
\texttt{TRICERATOPS} \citep{2020ascl.soft02004G, giacalone_vetting_2021}.
}

\bibliography{references, AdditionalReferences}

\end{document}